\documentclass[a4paper,11p]{article}
\pdfoutput=1 
\usepackage{jcappub} 

\usepackage[dvipsnames]{xcolor}

\usepackage{amsmath,amsfonts,amssymb,mathtools}
\allowdisplaybreaks 

\newcommand{\be}{\begin{equation}}
\newcommand{\ee}{\end{equation}}
\newcommand{\bea}{\begin{eqnarray}}
\newcommand{\eea}{\end{eqnarray}}
\newcommand{\bdm}{\begin{displaymath}}
\newcommand{\edm}{\end{displaymath}}
\newcommand{\nn}{\nonumber}

\newcommand{\xv}{\mathbf{x}}

\newcommand{\kv}{\mathbf{k}}

\newcommand{\qv}{\mathbf{q}}
\newcommand{\pv}{\mathbf{p}}
\newcommand{\sv}{\mathbf{s}}

\newcommand{\vv}{\mathbf{v}}

\newcommand{\del}{\delta}

\newcommand{\Le}{{\mathcal L}}
\newcommand{\kmax}{k_{\rm max}}

\newcommand{\kmaxB}{k_{\rm max,B}}

\newcommand{\bG}{b_{\mathcal G_2}}

\newcommand{\I}{\mathcal{I}}
\newcommand{\J}{\mathcal{J}}

\newcommand{\Mpc}{\, h^{-1} \, {\rm Mpc}}

\newcommand{\cGpc}{\, h^{-3} \, {\rm Gpc}^3}
\newcommand{\kMpc}{\, h \, {\rm Mpc}^{-1}}

\newcommand{\bGt}{b_{\mathcal{G}_2}}

\def\pin{{\sc{Pinocchio}}}

\defcitealias{OddoEtal2020}{Paper I}
\newcommand{\PaperI}{\citetalias{OddoEtal2020}}
\defcitealias{OddoEtal2021}{Paper II}
\newcommand{\PaperII}{\citetalias{OddoEtal2021}}

\title{The Halo Bispectrum Multipoles in Redshift Space}

\author[h,b,c,a,1]{Federico Rizzo,\note{Corresponding author.}}
\author[d,c]{Chiara Moretti,}
\author[a,e,f,c,h]{Kevin Pardede,}
\author[g,i]{Alexander Eggemeier,}
\author[a]{Andrea Oddo,}
\author[a,c,h]{Emiliano Sefusatti,}
\author[i]{Cristiano Porciani,}
\author[a,b,c]{and Pierluigi Monaco}

\affiliation[a]{Institute for Fundamental Physics of the Universe, Via Beirut 2, 34151 Trieste, Italy}
\affiliation[b]{Dipartimento di Fisica, Sezione di Astronomia, Universit\`a di Trieste, via Tiepolo 11, 34143 Trieste, Italy}
\affiliation[c]{Istituto Nazionale di Astrofisica, Osservatorio Astronomico di Trieste, via Tiepolo 11, 34143 Trieste, Italy}
\affiliation[d]{Institute for Astronomy, The University of Edinburgh, Royal Observatory, Edinburgh EH9 3HJ, UK}
\affiliation[e]{SISSA - International School for Advanced Studies, Via Bonomea 265, 34136 Trieste,  Italy}
\affiliation[f]{ICTP, International Centre for Theoretical Physics, Strada Costiera 11, 34151, Trieste, Italy}
\affiliation[g]{Institute for Computational Cosmology, Department of Physics, Durham University, South Road, Durham DH1 3LE, United Kingdom}
\affiliation[h]{Istituto Nazionale di Fisica Nucleare, Sezione di Trieste,  via  Valerio  2,  34127 Trieste,  Italy}
\affiliation[i]{Argelander Institut f\"ur Astronomie der Universit\"at Bonn, Auf dem H\"ugel 71, 53121 Bonn, Germany}

\emailAdd{federico.rizzo@inaf.it}

\abstract{ 
We present the analysis of the halo bispectrum in redshift-space in terms of its multipoles, monopole, quadrupole and hexadecapole, measured from a large set of simulations. We fit such measurements with a tree-level model in perturbation theory that depends on linear and nonlinear bias parameters as well as on the growth rate $f$ of density fluctuations. The likelihood analysis takes advantage of a very large set of mock catalogs, enabling a robust estimation of the covariance properties for all multipoles. We compare the numerical estimate of the covariance matrix to its Gaussian prediction finding discrepancies of 10\% or less for all configurations with the sole exception of the squeezed triangles in the monopole case. We find the range of validity of the tree-level model, for the total simulation volume of about 1000$\cGpc$, reaches a maximum wavenumber of $0.08\kMpc$ for the monopole, while it is limited to $0.06$ and $0.045\kMpc$ respectively for quadrupole and hexadecapole. Despite this, the addition of the quadrupole to the analysis allows for significant improvements on the determination of the model parameters and specifically on $f$, similarly to the power spectrum case. Finally, we compare our numerical estimate for the full covariance with its theoretical prediction in the Gaussian approximation and find the latter to work remarkably well in the context of simulation boxes with periodic boundary condition.       
}


\begin{document}
\maketitle

\section{Introduction}

As the next generation of spectroscopic galaxy surveys \cite{LaureijsEtal2011, LeviEtal2013, DoreEtal2014A} will cover unprecedented cosmological volumes, increasing attention is currently being payed to the full exploitation of the information they are expected to provide. Recent measurements and analyses of higher-order statistics such as the 3-point correlation function in configuration space \cite{SlepianEtal2017, VeropalumboEtal2021} or its counterpart in Fourier space, the bispectrum \cite{GilMarinEtal2017, PearsonSamushia2018, DAmicoEtal2020, PhilcoxIvanov2022, DAmicoEtal2022A, CabassEtal2022A, CabassEtal2022B}, go precisely in this direction, with the goal of extending and strengthening the results from the standard analyses of the 2-point correlation function and power spectrum. 

However, while the analysis of the power spectrum takes full advantage of redshift-space distortions by means of a multipoles expansion with respect to the angle between the wavenumber $\kv$ and the line-of-sight (see, e.g. \cite{AlamEtal2017, RuggeriEtal2019}), in the case of the bispectrum past data analyses have always been limited to the monopole. On the other hand, the potential offered by the galaxy bispectrum measured in future surveys to further constrain cosmological parameters has been explored  in several papers \cite{SongTaruyaOka2015, GagraniSamushia2017, YankelevichPorciani2019, ChudaykinIvanov2019, HahnEtal2020, GualdiVerde2020, HahnVillaescusaNavarro2021, AgarwalEtal2021, IvanovEtal2021A}. A subset of these works specifically considered the relevance of the anisotropic bispectrum signal \cite{SongTaruyaOka2015, GagraniSamushia2017, YankelevichPorciani2019, GualdiVerde2020, AgarwalEtal2021} remarking that we can expect additional information in the higher-order multipoles of the bispectrum, although the exact extent of such improvement on parameters constraints, typically of the order of tens of percents, highly depends on the assumptions on the observable, its covariance and the survey specifications.  

A first theoretical modelling of the redshift-space bispectrum at tree-level in Perturbation Theory can be found in \cite{HivonEtal1995} (see \cite{BernardeauEtal2002} and references therein for earlier work on the matter and galaxy bispectrum in real space). Early comparisons against measurements of the bispectrum monopole in numerical simulations are presented in \cite{VerdeEtal1998, ScoccimarroCouchmanFrieman1999, Scoccimarro2000B} with \cite{ScoccimarroCouchmanFrieman1999} including as well a first test of the quadrupole. 
The analysis of the BOSS data-set of \cite{GilMarinEtal2015, GilMarinEtal2015B, GilMarinEtal2017} includes the bispectrum monopole and takes advantage of a phenomenological model \cite{GilMarinEtal2014}, based on fits to simulations, to extend the validity of the tree-level expression to smaller scales, reaching $0.15\kMpc$ with a 5\% accuracy on the halo bispectrum monopole at redshift $z=0.55$ (to contrast $0.06\kMpc$ in the case of tree-level PT).  A similar approach is adopted as well in \cite{GualdiGilMarinVerde2021} where the monopole and quadrupole of the power spectrum, bispectrum and integrated trispectrum are compared to simulations. 

Ref. \cite{HashimotoRaseraTaruya2017} goes beyond the tree-level expression presenting a one-loop PT model for the redshift-space {\em matter} bispectrum multipoles (but defined differently from \cite{ScoccimarroCouchmanFrieman1999}), including additional corrections along the lines of those introduced by \cite{TaruyaNishimichiSaito2010} for the power spectrum. 
The comparison with numerical simulations shows an agreement up to $k\sim 0.15$ - $0.2\kMpc$, with the maximum range of this agreement depending on the redshift of the sample and on the shape of the specific triangular configuration considered. On the other hand, the corresponding tree-level approximation typically fails already around $k\sim 0.07$ - $0.08\kMpc$ for both the monopole and the quadrupole.

More recently, \cite{DAmicoEtal2020} re-analysed the BOSS bispectrum monopole adopting a tree-level model up to $0.1\kMpc$, although no comparison with simulations or details on model validation are provided. A further analysis, extending the model to include one-loop corrections and corrections due to primordial non-Gaussianity is presented in \cite{DAmicoEtal2022A}. A comparison with large-volume simulations can be found instead, again for the monopole only, in \cite{IvanovEtal2021A} for measurements obtained from the very large simulation set already adopted for the challenge paper \cite{NishimichiEtal2020}, corresponding to a cumulative volume of 566$\cGpc$: in this case as well the reach of the tree-level expression is found to be $k_{\rm max}\sim 0.08\kMpc$. The same pipeline for the bispectrum monopole analysis is applied to the BOSS data in \cite{PhilcoxIvanov2022, CabassEtal2022A, CabassEtal2022B}. 

It appears that, despite the recent attention, tests of the redshift-space galaxy bispectrum model have been rather limited. In fact, the current literature is for the most part focused on the bispectrum monopole with only partial assessments of higher-order multipoles predictions in PT. 

The main goal of this paper is to provide a rigorous and extensive comparison of the tree-level predictions for the halo bispectrum monopole, quadrupole and hexadecapole (as defined in \cite{ScoccimarroCouchmanFrieman1999}) against measurements in a very large set of numerical simulations ($\sim 1,000\cGpc$) while taking advantage of a robust estimate of their covariance properties from an even larger set of mock catalogs. Our work constitutes the natural continuation of a series of papers exploring in details the challenges of a joint analysis of the galaxy power spectrum and bispectrum, so far focused on real-space modelling \cite{OddoEtal2020, AlkhanishviliEtal2021, OddoEtal2021}. 
Since this work shares with these references both data-sets and methodology, we will refer to \cite{OddoEtal2020} and \cite{OddoEtal2021} as \PaperI{} and \PaperII{} respectively.
We test the model by means of a Bayesian analysis in terms of bias parameters along with the growth rate of perturbations $f$, using the simulation input and real-space results as reference values. 
The measurement uncertainties are reduced due to the large combined volume of our simulations. For such small errors, we find that at $z=1$ the model provides a valid description up to a maximum wavenumber of $0.08\kMpc$ for the monopole, $0.06\kMpc$ for the quadrupole, and $0.045\kMpc$ for the hexadecapole. 
We show that, as in the power spectrum case, the inclusion of the bispectrum quadrupole greatly improves the posteriors from the monopole alone.

The paper is organised as follows. In section 2 we introduce the theoretical background for the tree-level prediction of the bispectrum multipoles in Perturbation Theory. Section 3 describes the numerical simulations and mock catalogs adopted as well as the bispectrum estimator. In Section 4 we present the set-up for our likelihood analyses and in Section 5 the corresponding results. We present our conclusions in Section 6.

\section{Theoretical background}

\subsection{Model}
\label{ssec:model}

Given the halo number density contrast $\delta_h(\xv)\equiv [n_h(\xv)-\bar{n}_h]/\bar{n}_h$ defined in terms of the number density $n_h(\xv)$ and its expectation value $\bar{n}=\langle n_h(\xv)\rangle$, and its Fourier transform\footnote{We adopt the convention for the Fourier transform 
\be
\delta(\kv)  \equiv  \int d^3x\, e^{-i\kv\cdot\xv}\,\delta(\xv)\,,
\ee
with the inverse given by
\be
\delta(\xv)  \equiv  \int \frac{d^3k}{(2\pi)^3}\,e^{i\kv\cdot\xv}\,\delta(\kv)\,.
\ee} $\delta_h(\kv)$ we can define the halo power spectrum $P_h$ and bispectrum $B_h$ respectively as
\bea
\label{eq:Def_Ph}
\langle\delta_h(\kv_1)\delta_h(\kv_2)\rangle & \equiv & (2\pi)^3\delta_D(\kv_{12})\,P_h(k_1)\,\\
\label{eq:Def_Bh}
\langle\delta_h(\kv_1)\delta_h(\kv_2)\delta_h(\kv_3)\rangle & \equiv & (2\pi)^3\delta_D(\kv_{123})\,B_h(k_1,k_2,k_3)\,,
\eea
where $\kv_{12}=\kv_1+\kv_2$, $\kv_{123}=\kv_1+\kv_2+\kv_3$, and the Dirac deltas $\delta_D$ result from the assumed statistical homogeneity and isotropy. For the same reason $P_h(k_1)$ is a function of one variable, $k_1=|\kv_1|$ and $B_h(k_1,k_2,k_3)$ is a function of the three sides of the triangle formed by $\kv_1$, $\kv_2$ and $\kv_3$ and independent of its orientation.      

In redshift-space, peculiar velocities $\vv$ induce distortions in the galaxy distribution along the line-of-sight (LOS) $\hat{n}$. The observed position $\sv$ will then be related to real position $\xv$ by 
\be
\sv=\xv+\frac{\vv\cdot\hat{n}}{a\,H(a)}\hat{n}\,.
\ee
As a result, clustering properties, and in particular galaxy correlation functions estimated in a given region of the sky, will depend on the local LOS. Since our focus is to test the modelling of the bispectrum based on measurements in simulation boxes with periodic boundary conditions, we will assume throughout this work the plane-parallel approximation for redshift-space distortions and therefore a global, constant LOS. The halo bispectrum will then be a function of the wavenumbers defining the triangular configuration $\kv_1$, $\kv_2$ and $\kv_3$ plus the LOS  $\hat{n}$, that is $B_s=B_s( \textbf{k}_1, \textbf{k}_2, \hat{n})$. 

Our model for the redshift-space halo bispectrum is the sum of a deterministic and stochastic contribution, as
\be
\label{eq:Bmodel}
B_s(\kv_1,\kv_2,\kv_3)=B_s^{\rm (det)}(\kv_1,\kv_2,\kv_3)+B_s^{\rm (stoch)}(\kv_1,\kv_2,\kv_3)\,,
\ee
corresponding to the tree-level expression in Perturbation Theory (PT) resulting from the halo density given, in turn, by the sum of a deterministic and a stochastic component 
\be
\delta_s=\delta_s^{\rm (det)}+\delta_s^{\rm (stoch)}\,.
\ee
In Fourier space and up to the relevant order the deterministic contributions are given by
\be
\label{e:deltas_det}
\delta_s^{\rm (det)}(\kv)=Z_1(\kv)\,\delta_L(\kv)+\int d^3q_1 d^3q_2 \delta_D(\kv-\qv_{12})Z_2(\qv_1,\qv_2)\,\delta_L(\qv_1)\,\delta_L(\qv_1)\,,
\ee
where $\delta_L$ is the linear matter overdensity and the redshift-space kernels are given in terms of the local ($b_1$, $b_2$) and tidal ($\bGt$) bias parameters and the linear growth rate $f$ by \cite{FryGaztanaga1993, HivonEtal1995, VerdeEtal1998, ScoccimarroCouchmanFrieman1999, ChanScoccimarroSheth2012, BaldaufEtal2012, DesjacquesJeongSchmidt2018B}
\bea
Z_1(\kv) & = & b_1+f\mu^2\,,\\
Z_2(\kv_1, \kv_2) & = & \frac{b_2}{2}+b_1 F_2(\kv_1, \kv_2)+\bGt S(\kv_1, \kv_2)+f\mu_{12}^2G_2(\kv_1, \kv_2) +\nonumber\\
& & +\frac{f\mu_{12} k_{12}}{2}\left[\frac{\mu_1}{k_1}Z_1(\textbf{k}_2)+\frac{\mu_2}{k_2}Z_1(\textbf{k}_1)\right]
\eea
with $F_2$ and $G_2$ representing the usual matter density and velocity quadratic kernels and 
\be
S(\kv_1, \kv_2) = \left(\hat{k}_1\cdot\hat{k}_2\right)^2-1\,
\ee
while $\mu_i\equiv \kv_i\cdot\hat{n}/k_i$ is the cosine of the angle formed by the wavenumber $\kv_i$ with the LOS, specifically,
\be
\mu_{12}=\frac{\kv_{12}\cdot\hat{n}}{k_{12}} =-\frac{\textbf{k}_3\cdot\hat{n}}{k_3}=-\mu_3\,,
\ee
for a closed triangle with $\kv_{123}=0$. The expansion of eq.~(\ref{e:deltas_det}) leads to the tree-level prediction for the bispectrum
\be
\label{eq:B_det}
B_s^{\rm (det)}(\kv_1, \kv_2, \hat{n}) =  2\, Z_1(\kv_1)\,Z_1(\kv_2)Z_2(\kv_1, \kv_2)P_L(k_1)P_L(k_2)+{\rm 2~ perm.}\,
\ee
where $P_L(k)$ is the linear matter power spectrum.

The stochastic contribution to $\delta_s$ is given instead, following  \cite{DesjacquesJeongSchmidt2018B} and their notation, by
\be
\delta_s^{\rm (stoch)}(\xv)=\epsilon(\xv)+\epsilon_\delta(\xv)\,\delta(\xv)+ \epsilon_\eta(\xv)\eta(\xv)\,,
\ee
where $\epsilon$, $\epsilon_\delta$ and $\epsilon_\eta$ are stochastic fields uncorrelated to the density perturbations. 
The composite terms are limited to those linear in the matter density $\delta$ and in the l.o.s. derivative of the velocity component projected on the $\hat{n}$-axis $\eta\equiv\partial_{\hat{n}}(\vv\cdot \hat{n})$, as these are responsible for the leading order contributions to the bispectrum. 
We neglect any higher-derivative operator in the stochastic contribution and we note that the last term should appear only due to selection effects \cite{DesjacquesJeongSchmidt2018B}. 
In the large $k$ limit, we expect to recover the Poisson predictions for the power spectrum and bispectrum of the stochastic fields, that is \cite{Schmidt2016}
\begin{align}
\label{e:epsilon_limit_1}
\langle\epsilon(\kv_1)\epsilon(\kv_2)\rangle & \rightarrow \delta_D(\kv_{12})\,\frac{1}{\bar{n}}\,, \\
\label{e:epsilon_limit_2}
\langle\epsilon(\kv_1)\epsilon(\kv_2)\epsilon(\kv_3)\rangle & \rightarrow \delta_D(\kv_{123}) \, \frac{1}{\bar{n}^2}\,, \\
\langle\epsilon(\kv_1)\epsilon_\delta(\kv_2)\rangle & \rightarrow \delta_D(\kv_{123})\, \frac{b_1}{2\bar{n}}\,,  \\
\langle\epsilon(\kv_1)\epsilon_\eta(\kv_2)\rangle & \rightarrow \delta_D(\kv_{123})\, \frac{1}{2\bar{n}}\,,
\end{align}
where the first term only appears in the halo power spectrum, while the last three all contribute to the halo bispectrum. 
In principle we can expect independent departures from the Poisson prediction for all three terms, which in the large-scale limit can be described in terms of three constant parameters\footnote{In \cite{IvanovEtal2021A} the authors follow \cite{PerkoEtal2016A} in the modelling of the stochastic contribution assuming 
\be
\delta_s^{(stoch)}=d_1\,\epsilon_P +d_2\,b_1\,\epsilon_P\,\delta+d_1\,\epsilon_P\,\eta\,,
\ee
where the coefficients $d_1$ and $d_2$ parameterize the corrections to the Poisson prediction represented by field $\epsilon_P$ (for which the limits (\ref{e:epsilon_limit_1}) and (\ref{e:epsilon_limit_2}) hold as equalities). The Poisson case is recovered for $d_1=2\,d_2=1$. This implies that $\langle\epsilon\epsilon\rangle=\langle\epsilon\epsilon_\eta\rangle$ and their corrections to Poisson are therefore described by a single degree of freedom. They also relate $\langle\epsilon\epsilon\rangle$ and $\langle\epsilon\epsilon\epsilon\rangle$ but it does not seem justified. Such relation also appears inconsistent with the expansion above and it does not seem to be supported by the halo model description of \cite{GinzburgDesjacquesChan2017}.}.

The corresponding stochastic contribution to the bispectrum at tree-level will then read
\begin{align}
\label{eq:B_stochastic}
B_s^{\rm (stoch)}(\kv_1, \kv_2, \hat{n})  =  \frac{1}{\bar{n}}\left[(1+\alpha_1)\,b_1+(1+\alpha_3)\,f\,\mu^2\right]\,Z_1(\kv_1)\,P_L(k_1)+2~{\rm perm.}
+\frac{1+\alpha_2}{\bar{n}^2}\,,
\end{align}
where the parameters $\alpha_i$ vanish in the Poisson limit\footnote{The notation for the $\alpha_i$ parameters is chosen in order to be consistent with \PaperI{} and \PaperII{}, where $\alpha_2$ already appeared as correction to the $1/\bar{n}^2$ term, while $\alpha_3$ was not present.}. 

In this work we do not consider any modelling of Finger-of-God effects as we expect them to be negligible at large scales and for a halo distribution.

\subsection{Bispectrum multipoles}

We adopt the definition of the redshift-space multipoles of the bispectrum introduced by \cite{ScoccimarroCouchmanFrieman1999} (and assumed as well by \cite{Scoccimarro2015} and \cite{GagraniSamushia2017}) where the vector configurations covering the domain of $B_s(\kv_1\kv_2,\hat{n})$ are given in terms of the variables $k_1$, $k_2$, $k_3$, $\mu_1\equiv \cos(\theta)$ and $\xi$, with $\theta_1$ being the angle between $\kv_1$ and the LOS while $\xi$ is the azimuthal angle describing a rotation of $\kv_2$ around $\kv_1$.

$B_s$ is then expanded in spherical harmonics as
\be
\label{e:Bexp}
B_s(k_1, k_2, k_3, \theta, \xi) = \sum_{\ell}\sum_{m=-\ell}^{\ell} B_{\ell m}(k_1, k_2, k_3)\,Y_{\ell}^m(\theta, \xi)
\ee
with the coefficients of the expansion given by
\be
\label{e:Blm}
B_{\ell m}(k_1, k_2, k_3)=\int_{-1}^{+1}\!\!d\!\cos{\theta} \int_0^{2\pi}d\xi B_s(k_1, k_2, k_3, \theta,\xi)\,Y_{\ell}^m(\theta,\xi)\,.
\ee

We only consider $m=0$, even multipoles as the loss of information coming from excluding the $m\ne 0$ terms is negligible \cite{GagraniSamushia2017, ByunKrause2022}. In this case the spherical harmonics reduce to Legendre polynomials $\mathcal{L}_\ell$ and only depend on $\mu_1\equiv \cos\theta$, 
\be
Y_{\ell}^0(\theta,\xi)=\sqrt{\frac{2\ell +1}{4\pi}}\mathcal{L}_{\ell}(\mu_1)\,
\ee
and the expansion of eq.~(\ref{e:Bexp}) is replaced by
\be
\label{e:Bexp0}
\frac1{2\pi} \int d\xi B_s(k_1, k_2, k_3, \theta, \xi) = \sum_{\ell}\,B_{\ell}(k_1, k_2, k_3)\,\Le_\ell(\mu_1)
\ee
where
\begin{align}
\label{e:Bl}
B_\ell(k_1,k_2,k_3) & = \sqrt{\frac{2\ell+1}{4\pi}}\,B_{\ell 0}(k_1,k_2,k_3) \nn\\
& = (2\ell+1)\frac{1}{2} \int_{-1}^{+1}\!\!d\!\cos{\theta} \Big[\frac{1}{2\pi}\int_0^{2\pi}d\xi B_s(k_1, k_2, k_3, \theta,\xi)\Big]\,{\mathcal L}_{\ell}(\cos \theta)\,.
\end{align}

\section{Data}

\subsection{N-body simulations}

The analysis is performed on redshift-space, halo bispectrum measurements from the set of 298 Minerva \textit{N}-body simulations \cite{GriebEtal2016} whose real-space counterpart was already studied in \PaperI{} and \PaperII{}. These follow the evolution of $1000^3$ dark matter particles in a cubic box of side  $L=1500\Mpc$ and correspond to a total volume of about $1,000\cGpc$. Each halo catalog is defined by a minimal mass of $M\simeq 1.12\times 10^{13}\; h^{-1} M_{\odot}$. We refer the reader to  \PaperI{} for a more detailed description of the simulations and of the halo catalog construction.

\PaperII{} provides us with an estimate of the bias parameters characterising the halo population obtained as posteriors from the joint analysis of the halo power spectrum and bispectrum in real space. We will use these here as a reference for our redshift-space analysis, in addition to the value for the linear growth rate expected from the fiducial cosmology.

\subsection{Bispectrum multipoles estimator}
\label{ssec:estimators}

Our estimator of the bispectrum multipoles follow the definition of \cite{ScoccimarroCouchmanFrieman1999} for the $m=0$ case and constitutes an implementation of the one described in \cite{Scoccimarro2015} based on Fast-Fourier Transforms. In our case, however, we assume a constant line-of-sight $\hat{n}$, corresponding to an exact realisation of the plane-parallel or distant observer approximation. The estimator reduces therefore to the following expression
\be
\label{e:Best}
\hat{B}_\ell = (2\ell +1)\frac{k_f^3}{N_B} \sum_{\qv_1\in k_1} \sum_{\qv_2\in k_2} \sum_{\qv_3\in k_3} \,\delta_K(\qv_{123})\,\delta_s(\qv_1)\,\delta_s(\qv_2)\,\delta_s(\qv_3)\,\mathcal{L}_\ell(\hat{q}_1\cdot\hat{n})\,,
\ee
where the sums, accounting for the discrete nature of the Fourier Transform $\delta_s(\qv)$ of the halo density in a simulation box, are over all wavenumbers $\qv_i$ falling into the bin centered at $k_i$ of radial size $\Delta k$, that is, such that $k_i-\Delta k/2\le |\qv_i|<k_i+\Delta k/2$. Also, $\delta_K(\qv)$ is a Kronecker symbol equal to unity for $\qv={\bf 0}$ and vanishing otherwise, while the normalisation factor
\be
\label{e:NB}
N_B(k_1,k_2,k_3) =  \sum_{\qv_1\in k_1} \sum_{\qv_2\in k_2} \sum_{\qv_3\in k_3} \,\delta_K(\qv_{123})\,,
\ee
provides the number of ``fundamental triangles'' $\{\qv_1, \qv_2, \qv_3\}$ present in the ``triangle bin'' $\{k_1,k_2,k_3\}$.  
The grid-interpolation of the halo density $\delta_s(\qv)$ is obtained by means of a fourth-order mass assignment scheme and adopts the interlacing technique for aliasing reduction \cite{SefusattiEtal2016}.  
All bispectrum measurements, unless otherwise stated, assume a wavenumber bin size $\Delta k= k_f$, that is corresponding to the fundamental frequency characterising the simulation box, $k_f\equiv 2\pi/L$. This leads to the measurement of 1475 triangular configurations up to $\kmax=0.1\kMpc$ for each multipole\footnote{We include ``open triangle bins'', that is those where the bin centers cannot form a closed triangle such as $\{k_1,k_2,k_3\}=\{6,3,2\}k_f$ but that nevertheless contain closed fundamental triplets $\{\qv_1, \qv_2, \qv_3\}$. See section 2.2 of \PaperI{} for a detailed description of the binning definition.}.

\subsection{Measurements}

\begin{figure}[t]
    \centering
    \includegraphics[width=1.\textwidth]{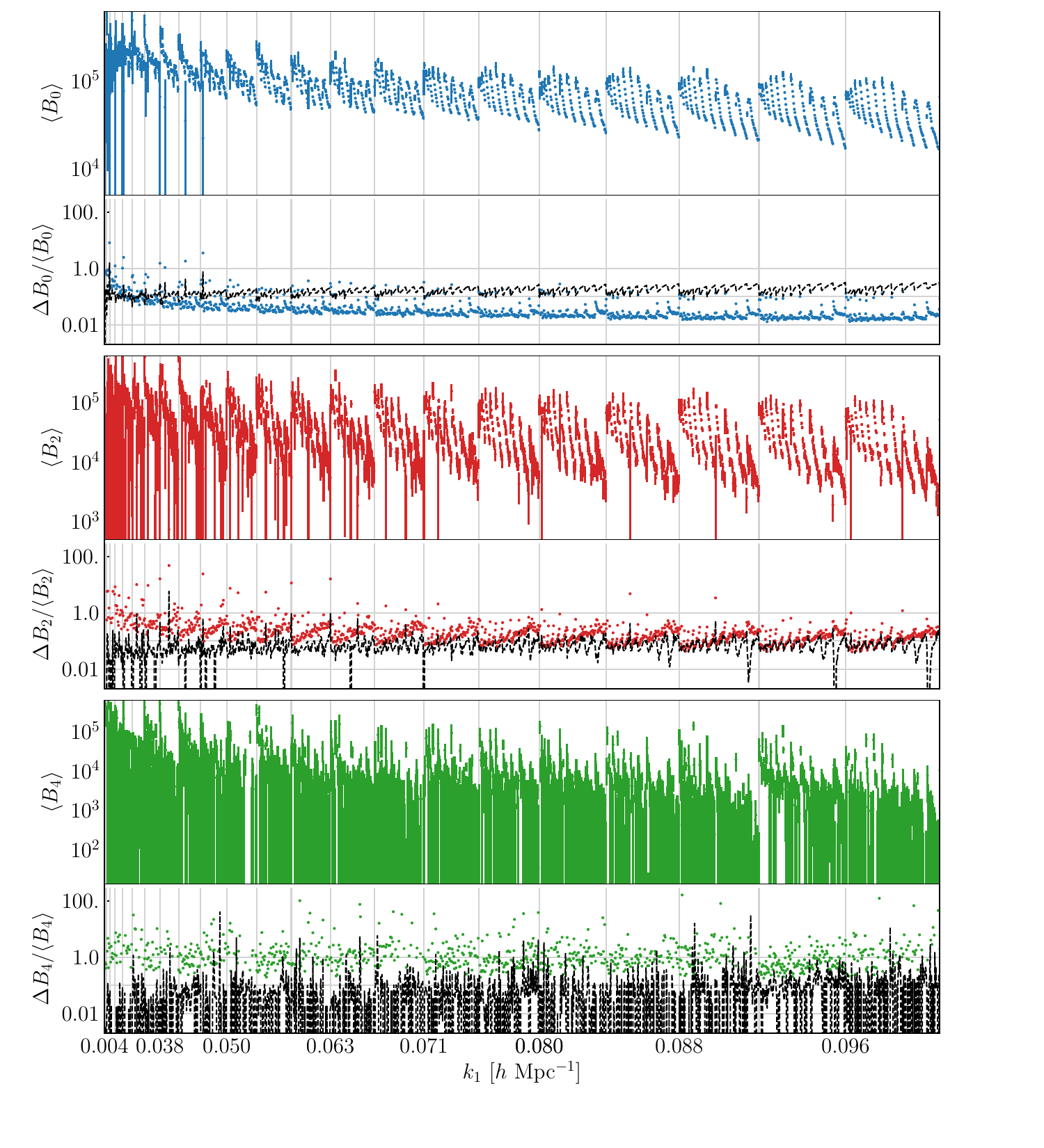}
    \caption{Mean measurements of the halo bispectrum multipoles (including shot-noise) from the 298 N-body Minerva simulations, shown for all triangular configurations. For each multipole moment, the bottom half of each panel shows the relative error of the mean (in color) along with the relative (Poisson) shot-noise contribution (in black).}
    \label{fig:measurements_all}
\end{figure}

Figure \ref{fig:measurements_all} shows the mean of the bispectrum multipoles measured from the 298 Minerva N-body simulations for all triangular configurations. In these type of plots, the ordering of the configurations is determined by increasing values of $k_1$, $k_2$, $k_3$ which obey the requirement, $k_1\ge k_2\ge k_3$ (see \PaperI{} for a more detailed explanation). Vertical gray lines mark the triangle where the value of $k_1$ changes, so that all configurations on the left correspond to triangles made up with sides smaller or equal to such value of $k_1$. All measurements include shot-noise. The bottom half of each panel shows the relative error on the mean, along with the ratio between the expected Poisson shot-noise contribution and the overall signal (black, dashed lines).    

We notice that the relative error on the mean for the bispectrum monopole is at the ten-percent level and just slightly smaller at smaller scales. The shot-noise level is comparable to the statistical error at large scales and it is larger at smaller scales, as it happens in real-space for this halo population (see \PaperI). The relative error on the quadrupole and hexadecapole mean is instead of the order of tens of percent and hundreds of percent, respectively. Because of the first term on the r.h.s. of \eqref{eq:B_stochastic}, all multipoles receive a shot-noise contribution. This is comparable to the error on the mean for the quadrupole and lower by an order of magnitude for the hexadecapole.

\section{Covariance}
\label{ssec:covariance}

\subsection{Numerical estimate}

As for \PaperI{} and \PaperII, the covariance properties of our observables are estimated from a much larger set of 10,000 measurements from mock halo catalogs obtained with the \pin{} code \cite{MonacoTheunsTaffoni2002, MonacoEtal2013, MunariEtal2017}. The mocks share the same cosmology, box size and resolution with the Minerva simulations and 298 realisations also adopt the same initial conditions. The mass threshold for the mocks is chosen to reproduce (below percent level) the amplitude of the large-scale halo power spectrum (including shot-noise) of the numerical simulations (see \PaperI{} for details). This quantity, in fact, determines the Gaussian contribution to the bispectrum covariance, the leading one for most triangular configurations (see also \cite{Barreira2019, BiagettiEtal2021A}). 

The covariance matrix for each bispectrum multipole and their cross-covariance is defined as
\be
C_{\ell_1\ell_2}(t_i,t_j)\equiv \langle \hat{B}_{\ell_1}(t_i)\,\hat{B}_{\ell_2}(t_j)\rangle-\langle \hat{B}_{\ell_1}(t_i)\rangle\langle\hat{B}_{\ell_2}(t_j)\rangle\,,
\ee
where $t_i=\{k_{1i},k_{2i},k_{3i}\}$ and $t_j=\{k_{1j},k_{2j},k_{3j}\}$ represent two triangle configurations. We will denote as $\hat{C}_{\ell_1\ell_2}(t_i,t_j)$ its estimate from the 10,000 mock catalogs. 

\begin{figure}[t]
    \centering
    \includegraphics[width=1\textwidth]{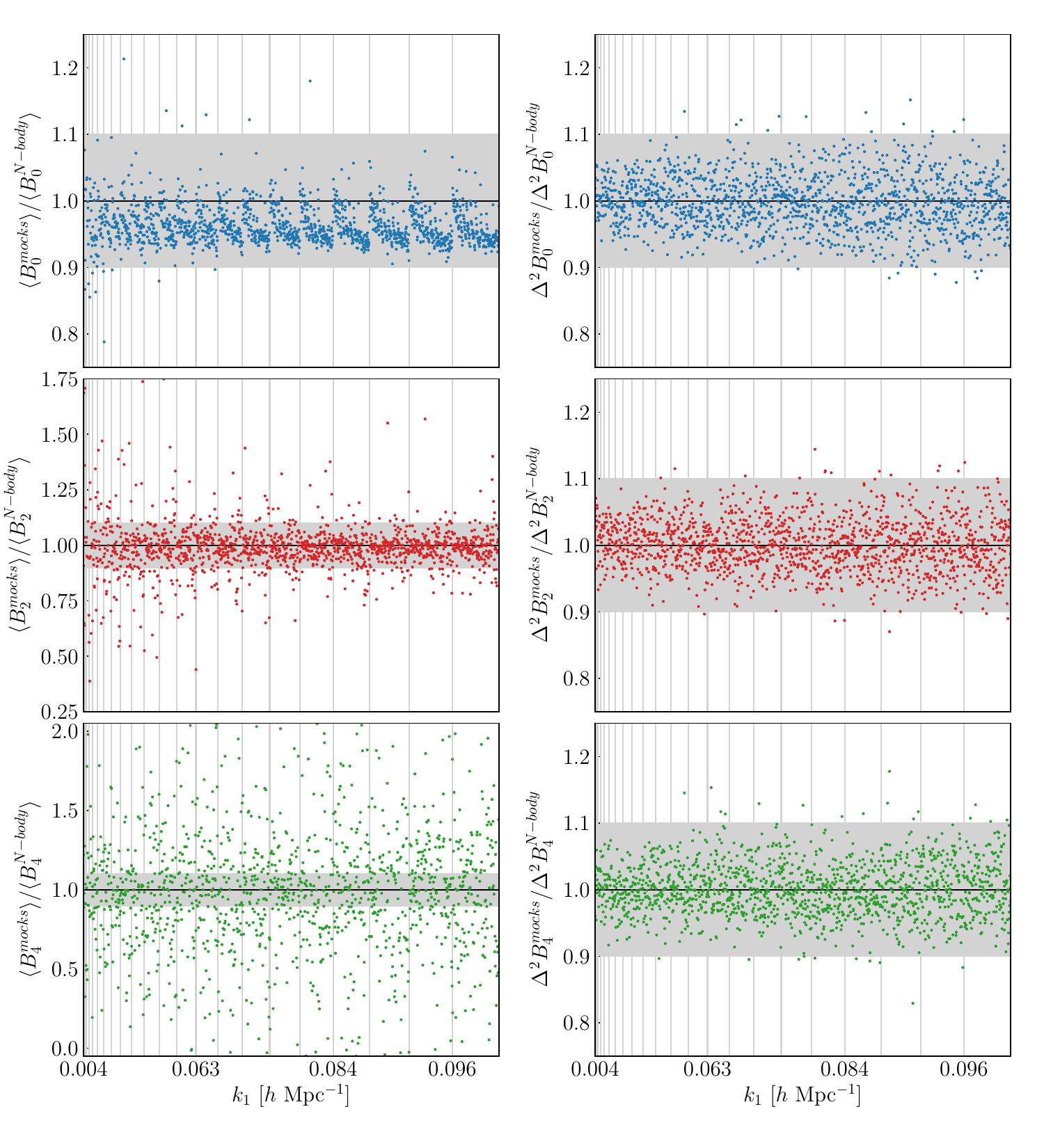}
    \caption{Left column: ratio between the mean bispectrum multipoles from the 298 N-body simulations and the mean of the same quantity from the corresponding \pin{} mocks with matching initial conditions. Right column: ratio between the bispectrum multipoles variance estimated from the simulations and the one estimated from the corresponding mocks.}
    \label{fig:variance}
\end{figure}

The left column of figure~\ref{fig:variance} shows the ratio between the mean of each bispectrum multipole measured in the numerical simulations and the mean of the same quantity measured in the \pin{} mocks, limited to the 298 mocks with matching initial conditions. We find for the monopole the same discrepancy, as large as 7-8\% depending on the triangle shape, already encountered in real space (see \PaperI). The noise in the measurements for the higher-order multipoles, on the other hand, does not allow to clearly identify systematic differences at the level of 10\% or below.    
The right column of figure~\ref{fig:variance} shows instead the ratio between the variance $\Delta B_\ell(t_i)\equiv C_{\ell\ell}(t_i,t_i)$, estimated again from the numerical simulations and the one estimated from the \pin{} mocks. Again, the \pin{} set is limited here to the 298 realisations with matching seeds. Despite possible systematics on the observables, the variance is recovered by the \pin{} mocks with an error below 10\% and no apparent systematic difference for all multipoles. This is expected, given the close match of the power spectra and the fact that the leading contribution to the bispectrum covariance is fully determined by the power spectrum, see Eq.~(\ref{eq:theory_cov.Gaussian}).

\begin{figure}[t]
    \centering
    \includegraphics[width=0.95\textwidth]{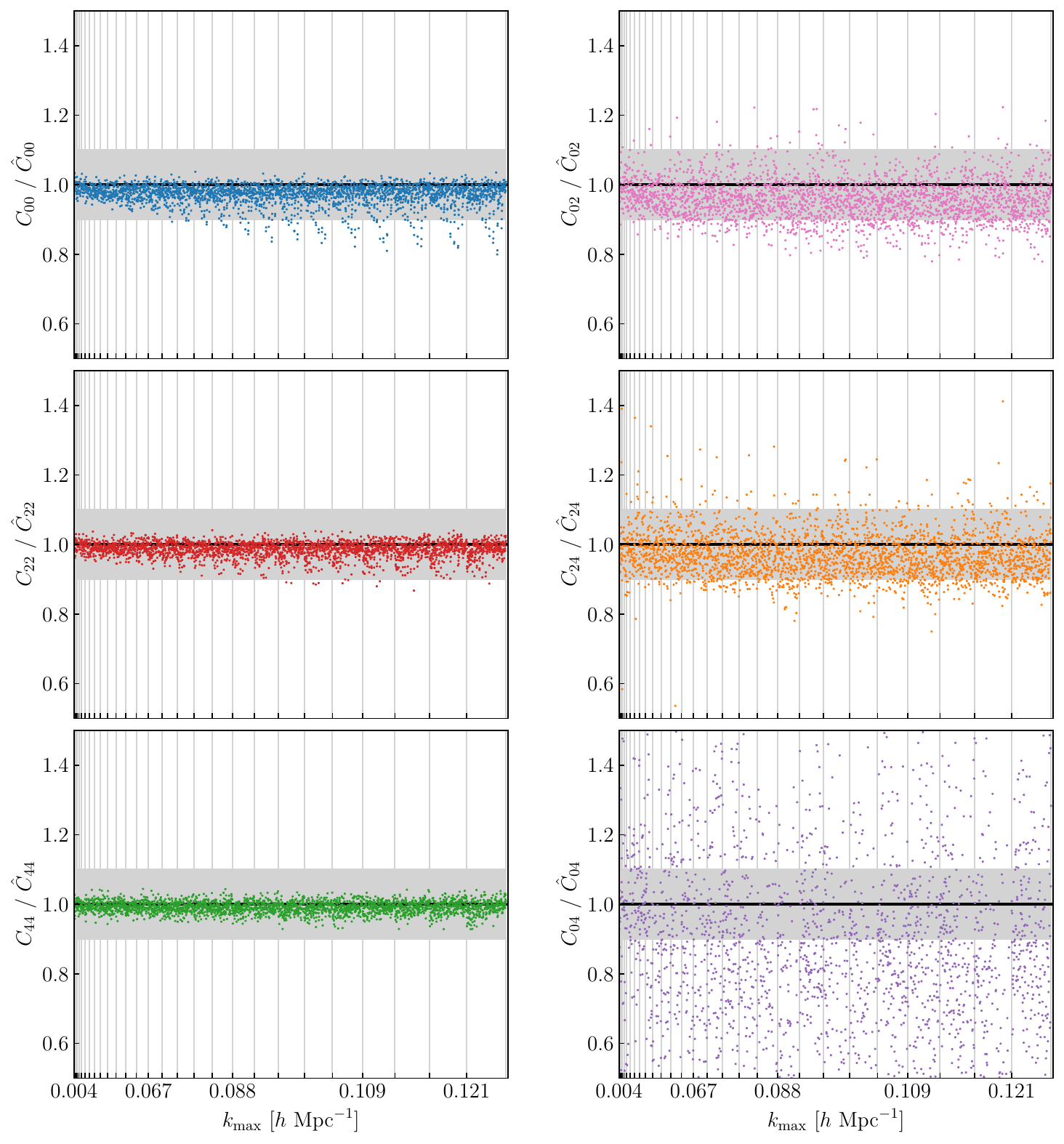}
    \caption{Ratio between the theoretical prediction of the multipoles covariance $C_{\ell_1\ell_2}(t_i,t_i)$ in the Gaussian approximation and its numerical estimate from the \pin{} mocks.}
    \label{fig:variance_theory}
\end{figure}

\subsection{Theoretical Gaussian covariance}

In addition to the numerical estimate of the covariance we consider as well the analytical prediction in the Gaussian approximation. The Gaussian contribution enters the variance of each multipole, $\Delta B_\ell\equiv C_{\ell\ell}(t_i,t_i)$, but also the correlation $C_{\ell_1\ell_2}(t_i,t_i)$ between $\hat{B}_{\ell_1}(t_i)$ and $\hat{B}_{\ell_2}(t_i)$ with $\ell_1\ne\ell_2$ but measured for the same triangle $t_i$.    

From the definition of the multipoles estimators, eq.~(\ref{e:Best}), we can write 
\begin{align}
C_{\ell_1\ell_2}(k_{1},k_{2},k_{3}) & \equiv  \langle \hat{B}_{\ell_1}\hat{B}_{\ell_2}\rangle -\langle \hat{B}_{\ell_1}\rangle\langle \hat{B}_{\ell_2}\rangle
\nn\\
& = (2\ell_1+1)(2\ell_2+1)\frac{k_f^6}{N^2_B} 
\sum_{\qv_1\in k_1}\sum_{\qv_2\in k_2}\sum_{\qv_3\in k_3}\,\del_K(\qv_{123})\,
\sum_{\pv_1\in k_1}\sum_{\pv_2\in k_2}\sum_{\pv_3\in k_3}\,\del_K(\pv_{123})
\nn \\ & \times 
\,\mathcal{L}_{\ell_1}(\mu_{\qv_1})\,\mathcal{L}_{\ell_2}(\mu_{\pv_1})\,\Big[\langle \delta_{\qv_1}\,\delta_{\qv_2}\,\delta_{\qv_3}\delta_{\pv_1}\,\delta_{\pv_2}\,\delta_{\pv_3}\rangle-
\langle \delta_{\qv_1}\,\delta_{\qv_2}\,\delta_{\qv_3}\rangle\langle\delta_{\pv_1}\,\delta_{\pv_2}\,\delta_{\pv_3}\rangle\Big]\,.
\end{align}
In the Gaussian approximation, from the expectation values on the r.h.s. of the equation above, we retain only the contributions depending on the power spectrum. Assuming (without loss of generality) that $k_1\ge k_2\ge k_3$, these are given by   
\begin{align}
\label{eq:theory_cov.Gaussian}
C_{\ell_1\ell_2}^G(k_{1},k_{2},k_{3}) 
& =  \frac{(2\ell_1+1)(2\ell_2+1)}{N^2_B\,k_f^3} \sum_{\qv_1\in k_1}\sum_{\qv_2\in k_2}\sum_{\qv_3\in k_3}\,\del_K(\qv_{123})\,P_{tot}(\qv_1)\,P_{tot}(\qv_2)\,P_{tot}(\qv_3)
\nn \\ & \times 
\Big[(1+\delta^K_{k_2,k_3})\,\mathcal{L}_{\ell_1}(\mu_{\qv_1})\mathcal{L}_{\ell_2}(-\mu_{\qv_1})+(\delta^K_{k_1,k_2}+\delta^K_{k_2,k_3})\,\mathcal{L}_{\ell_1}(\mu_{\qv_1})\mathcal{L}_{\ell_2}(-\mu_{\qv_2}) +
\nn \\ & +
2\,\delta^K_{k_1,k_3}\,\mathcal{L}_{\ell_1}(\mu_{\qv_1})\mathcal{L}_{\ell_2}(-\mu_{\qv_3})\Big]
\,,
\end{align}
where $P_{tot}(\qv_1)=P(\qv_1)+P_{SN}$ is the anisotropic halo power spectrum including a shot-noise contribution while $\delta^K_{k_i,k_j}$ is the Kronecker symbol equal to one for $k_i=k_j$, and vanishing otherwise.
Notice that the terms in the square brackets correspond, in the case $\ell_1=\ell_2=0$, to the usual factor equal to 6, 2 and 1 respectively for equilateral, isosceles and scalene triangles.

Similarly to the power spectrum variance case, see e.g. \cite{GriebEtal2016}, we can expand the anisotropic power spectra in multipoles to obtain
\begin{align}
C_{\ell_1\ell_2}^G 
& =  \frac{(2\ell_1+1)(2\ell_2+1)}{N^2_B\,k_f^3}\sum_{\ell_3,\ell_4,\ell_5} 
\sum_{\qv_1\in k_1}\sum_{\qv_2\in k_2}\sum_{\qv_3\in k_3}\,\del_K(\qv_{123})\,
P_{tot,\,\ell_3}(q_1)\,P_{tot,\,\ell_4}(q_2)\,P_{tot,\,\ell_5}(q_3)
\nn \\ & \times 
\Big[(1+\delta^K_{k_2,k_3})\,\mathcal{L}_{\ell_1}(\mu_{\qv_1})\mathcal{L}_{\ell_2}(-\mu_{\qv_1})+(\delta^K_{k_1,k_2}+\delta^K_{k_2,k_3})\,\mathcal{L}_{\ell_1}(\mu_{\qv_1})\mathcal{L}_{\ell_2}(-\mu_{\qv_2}) +
\nn \\ & +
2\,\delta^K_{k_1,k_3}\,\mathcal{L}_{\ell_1}(\mu_{\qv_1})\mathcal{L}_{\ell_2}(-\mu_{\qv_3})\Big]
\mathcal{L}_{\ell_3}(\mu_1)\mathcal{L}_{\ell_4}(\mu_2)\mathcal{L}_{\ell_5}(\mu_3)
\,.
\end{align}
This is the expression we adopt in our evaluation of the Gaussian variance, with the sums over the $k$-shells performed exactly over the discrete wavenumbers $\qv$ defining the Fourier-space density grid\footnote{It is possible to simplify further this expression in the thin-shell approximation so that
\begin{align}
C_{\ell_1\ell_2}^G 
& \simeq  \frac{(2\ell_1+1)(2\ell_2+1)}{N_B\,k_f^3}\sum_{\ell_3,\ell_4,\ell_5} 
P_{tot,\,\ell_3}(k_1)\,P_{tot,\,\ell_4}(k_2)\,P_{tot,\,\ell_5}(k_3)
R_{\ell_1,\ell_2;\ell_3\ell_4,\ell_5}(k_1,k_2,k_3)
\,.
\end{align}
where we defined
\begin{align}
R_{\ell_1,\ell_2;\ell_3\ell_4,\ell_5}(k_1,k_2,k_3)
& \equiv  \frac{1}{N_B}
\sum_{\qv_1\in k_1}\sum_{\qv_2\in k_2}\sum_{\qv_3\in k_3}\,\del_K(\qv_{123})\,
\nn \\ & \times 
\Big[(1+\delta^K_{k_2,k_3})\,\mathcal{L}_{\ell_1}(\mu_{\qv_1})\mathcal{L}_{\ell_2}(-\mu_{\qv_1})+(\delta^K_{k_1,k_2}+\delta^K_{k_2,k_3})\,\mathcal{L}_{\ell_1}(\mu_{\qv_1})\mathcal{L}_{\ell_2}(-\mu_{\qv_2}) +
\nn \\ & +
2\,\delta^K_{k_1,k_3}\,\mathcal{L}_{\ell_1}(\mu_{\qv_1})\mathcal{L}_{\ell_2}(-\mu_{\qv_3})\Big]
\mathcal{L}_{\ell_3}(\mu_1)\mathcal{L}_{\ell_4}(\mu_2)\mathcal{L}_{\ell_5}(\mu_3)
\,.
\end{align}
In the continuum limit, we can replace the sums over the shells with integrals and reduced them to a simple average over the orientation of the triangle $\left\{\qv_1,\qv_2,\qv_3\right\}$, that is
\be
 \frac{1}{N_B} \sum_{\qv_1\in k_1}\sum_{\qv_2\in k_2}\sum_{\qv_3\in k_3}\,\del_K(\qv_{123})\simeq \frac{1}{N_B\,k_f^6} \prod_{i=1}^3\int_{k_i}d^3q_i\,\del_D(\qv_{123}) = \frac{1}{4\pi}\int_{-1}^1d\mu_1\!\!\int d\xi\,,
\ee
with an integrand that is now only a function of powers of $\mu_1$, $\mu_2$ and $\mu_3$. Then, assumming only even values for $\ell_3$, $\ell_4$ and $\ell_5$ (and clearly for $\ell_1$ and $\ell_2$), we can use the expansion for the Legendre polynomials
\be
\mathcal{L}(\mu)=\frac{1}{2^\ell}\sum_{n=1}^{\ell/2}\frac{(-1)^n(2\ell-2n)!}{n!(\ell-n)!(\ell-2n)!}\mu^{\ell-2n}\equiv\sum_{n=1}^{\ell/2}C_{\ell,n}\mu^{\ell-2n}
\ee
to get an expression that can be automatically evaluated with a software allowing for symbolic manipulation,
\begin{align}
 R_{\ell_1,\ell_2;\ell_3\ell_4,\ell_5}(k_1,k_2,k_3)
& \simeq
 \prod_{i=1}^{5}\sum_{n_i=1}^{\ell_i/2}C_{\ell_i,n_i}\Big[(1+\delta^K_{k_2,k_3})\I_{
\ell_1+\ell_2+\ell_3-2(n_1+n_2+n_3),\,\ell_4-2n_4,\,\ell_5-2n_5}
\nn \\ & +
(\delta^K_{k_1,k_2}+\delta^K_{k_2,k_3})\I_{
\ell_1+\ell_3-2(n_1+n_3),\,\ell_2+\ell_4-2(n_2+n_4),\,\ell_5-2n_5}
\nn \\ & +
2\,\delta^K_{k_1,k_3}\,\I_{
\ell_1+\ell_3-2(n_1+n_3),\,\ell_4-2n_4,\,\ell_2+\ell_5-2(n_2+n_5)}\Big]
\,,
\end{align},
where the integrals $\I_{\alpha,\beta,\gamma}$ are defined in Appendix~\ref{app:A}.
}.

The comparison between the numerical estimate and analytic prediction for these quantities is shown in figure~\ref{fig:variance_theory}. One can see how the Gaussian prediction for $C_{00}$, $C_{22}$ and $C_{44}$ is able to describe the measured ones at the level of 5\%, with a slight deficit noticeable in the monopole and quadrupole case for squeezed triangles, due to the missing non-Gaussian contribution \cite{Barreira2019, BiagettiEtal2021A}. The agreement in the case of the cross-covariance $C_{\ell_1\ell_2}(t_i,t_i)$ is also rather good, with the theory underestimating the measurements by an overall 10\%. In these cases the ratio can take large values when the denominator is close to zero, as it is the case particularly for $C_{0,4}$.

\begin{figure}[t]
    \centering
    \includegraphics[width=0.89\textwidth]{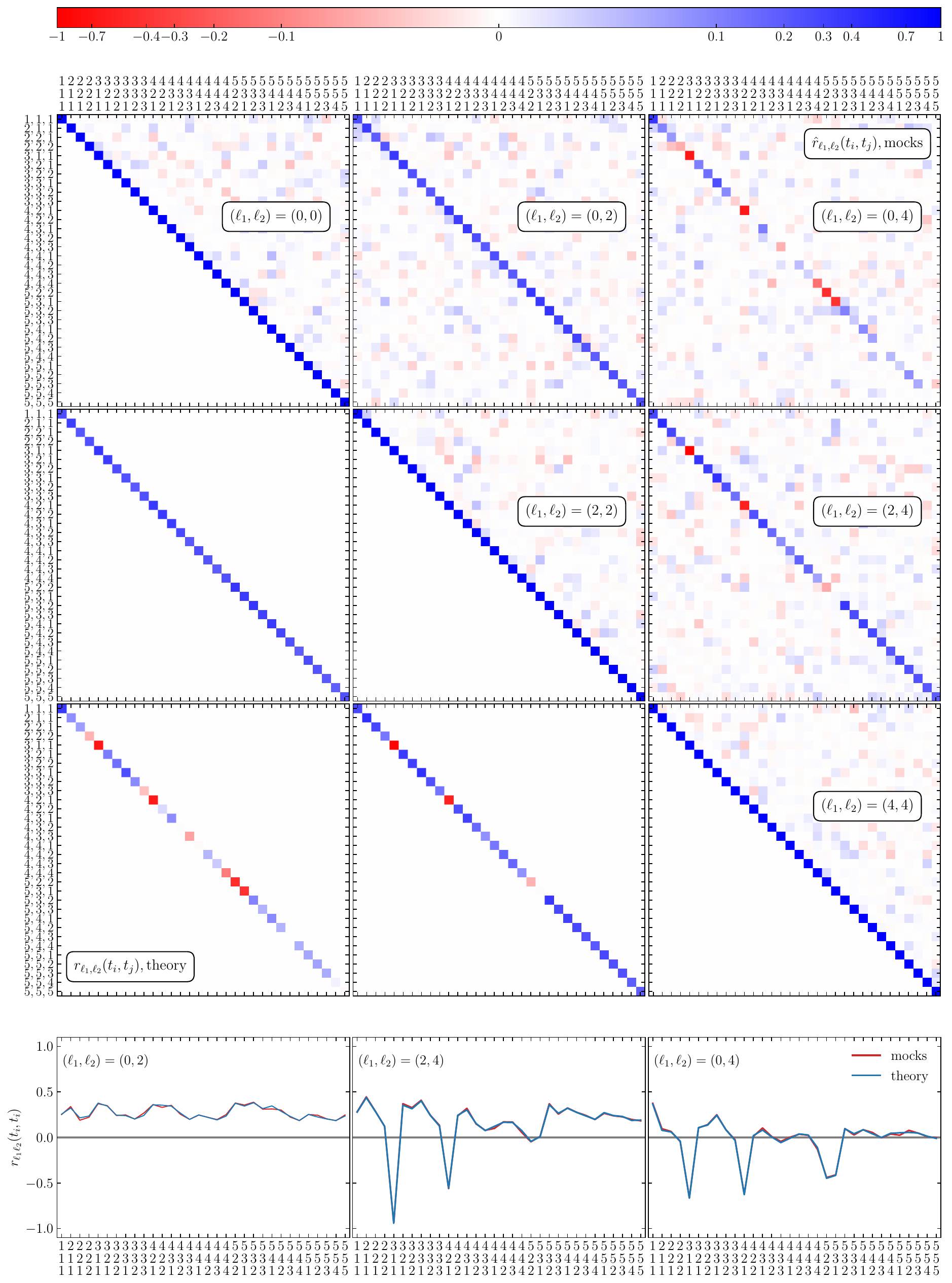}
    \caption{Subset of the correlation matrix $r_{\ell_1\ell_2}(t_i,t_j)$, defined in eq.~(\ref{eq:correlation_matrix}) for $\ell=0,2$ and 4, restricted to the first 32 triangular configurations $t_i$. The top-right half is estimated from the full set of 10,000 \pin{} mocks, while the bottom-left half is the theoretical prediction in the Gaussian approximation. The bottom panels compare the predicted (blue) and measured (red) coefficients $r_{\ell_1\ell_2}(t_i,t_i)$ with $\ell_1\ne\ell_2$ as a function of the selected triangles $t_i$. The two estimates overlap almost exactly.}
    \label{fig:cov}
\end{figure}

Finally, figure~\ref{fig:cov} shows a subset of the correlation matrix, defined as
\be
\label{eq:correlation_matrix}
r_{\ell_1\ell_2}(t_i,t_j)=\frac{C_{\ell_1\ell_2}(t_i,t_j)}{\sqrt{C_{\ell_1\ell_1}(t_i,t_i)\,C_{\ell_2\ell_2}(t_j,t_j)}}\,.
\ee
Each data-set $\hat{B}_\ell(t_i)$ is restricted, for illustration purposes, to its first 32 triangular configurations $t_i$, denoted in the figure in terms of the three sides in units of the fundamental frequency, that is $\{k_1,k_2,k_3\}/k_f$. It follows that while the block-diagonal matrices show the correlation coefficients $r_{\ell\ell}(t_i,t_j)$ for each multipoles, the off-diagonal matrices depict their relative cross-covariance $r_{\ell_1\ell_2}(t_i,t_j)$. In addition, the top-right half is estimated from the full set of 10,000 \pin{} mocks, while the bottom-left half is the theoretical prediction in the Gaussian approximation, vanishing for all elements with $t_i \ne t_j$. The bottom panels compare in more detail the predicted and measured coefficients $r_{\ell_1\ell_2}(t_i,t_i)$ with $\ell_1\ne\ell_2$ as a function of the selected triangles $t_i$. On these quantities the agreement between theory and numerical estimates is truly remarkable and extends up to $\kmax\sim 0.1\kMpc$, that is over the full range of scales that we will consider in the analysis described in the following sections.   

Regarding the structure of the correlation coefficient, it is clear that only the elements corresponding to the expected Gaussian contributions appear to be relevant at these large scales. These, however, are not limited to the diagonal for the full data vector $D=\left\{B_0,B_2,B_4\right\}$, but obviously include all elements corresponding to the correlation between different multipoles sharing the same triangles.

\section{Bayesian analysis}

\subsection{Likelihood function}

Following \PaperI{} and \PaperII{}, we fit all measurements together assuming their independence. This means that our total log-likelihood function corresponds to the sum of the log-likelihood for each individual realisation, 
\be
\ln {\mathcal{L}}_{\rm tot} = \sum_{\alpha=1}^{N_R}\ln {\mathcal L}_{\rm \alpha}\,,
\ee
where $N_R$ is the total number of realisations considered.    

We work under the assumption of Gaussianity for the individual likelihood ${\mathcal L}_{\rm \alpha}$. However, we follow \cite{SellentinHeavens2016} in order to account for possible uncertainties in the determination of the precision matrix due to a limited number of mocks. The log-likelihood for a single realisation is then, modulo an additive, normalization constant,
\be
\ln {\mathcal{L}}_{\alpha} = -\frac{N_M}2\ln \Big[1+\frac{\chi^2_{\alpha}}{N_M-1}\Big]\,,
\ee
where $N_M$ is the number of mock catalogs used for the numerical estimation of the covariance matrix (we refer the reader to \PaperI{} for further details). In this expression $\chi_\alpha^2$ represents the chi-square statistic for the individual realisation, given by
\be
\chi^2_{\alpha} = \sum_{i,j=1}^{N_D}\left[ \hat{D}^{(\alpha)}_{i}-D^{(theory)}_{i}\right] C_{ij}^{-1} \left[ \hat{D}^{(\alpha)}_{j}-D^{(theory)}_{j}\right]\,,
\ee
where, in the most general case, $\hat{D}^{(\alpha)}\equiv\left\{\hat{B}_0,\hat{B}_2,\hat{B}_4\right\}$ is the data vector, of size $N_D$, encompassing the three bispectrum multipoles while $D^{(theory)}$ and $C_{ij}$ are, respectively, the corresponding theoretical model and covariance matrix. 

We should notice that given the large number of 10,000 \pin{} realisations, even for the largest data-set corresponding to the joint analysis of the three bispectrum multipoles up to $\kmaxB=0.1\kMpc$, with a total of $1,475\times 3\simeq 4,425$ data-points, the difference w.r.t. the Gaussian case is in fact negligible. The alternative approach of re-scaling the inverse covariance, as suggested in \cite{Anderson2003, HartlapEtal2009}, gives rise to error bars up to 10\% larger, although we have checked that these do not lead to any appreciable differences in the recovered parameter posteriors that we discuss below.

\subsection{Model evaluation}

Our main goal is assessing the validity and reach of the tree-level bispectrum model, eqs.~(\ref{eq:B_det}) and (\ref{eq:B_stochastic}) leaving the exploration of their potential to constrain cosmological parameters to a future work. For this reason, in our Bayesian analyses we assume galaxy bias, shot-noise and the growth rate $f$ as the only free parameters. The bispectrum multipoles defined in eq.~(\ref{e:Blm}) can be written as a linear combination of several contributions where the dependence on these parameters can be factorised, leading to a quick exploration of the likelihood function since each term only needs to be computed once for the fiducial cosmology.

This allows as well for an exact binning of the theoretical model, taking advantage of the discrete Fourier-space grid characterising numerical simulations in boxes with periodic boundary conditions. In this case we can sum $B_s(\qv_1,\qv_2,\qv_3,\hat{n})$ over all discrete modes $\qv_i$ forming a close triangle $\qv_{123}=0$ and belonging to the bin $\{k_1,k_2,k_3\}$. This leads to     
\be
\label{e:exact_binning}
B_{\ell}^{\rm (binned)}(k_1, k_2, k_3) = \frac{2\ell+1}{N_B}\sum_{\qv_1 \in k_1}\sum_{\qv_2 \in k_2}\sum_{\qv_3 \in k_3} \delta_K(\qv_{123})B_{s}(\qv_1, \qv_2, \qv_3,\hat{n}) \mathcal{L}_{\ell}(\hat{q}_1\cdot\hat{n}),
\ee
where the sums account for the angle-average defining the bispectrum multipoles.

This approach requires the evaluation of the bispectrum model over a very large number of triangular configurations, making it unfeasible in a Bayesian analysis where cosmological parameters are explored. An approximate solution would be to evaluate the model $B_{\ell}$ on a single triangle defined by effective values of the wavenumbers. This approach, that takes advantage of the analytical evaluation of the angle integrals in \eqref{e:Bl} described in Appendix \ref{app:A}, is presented in Appendix \ref{app:EffectiveBinning} along with a quantification of the systematic errors resulting in the parameters determination.

\subsection{Goodness of fit and model selection}
\label{ssec:goodness}

We will assess the goodness of the fits that we will perform in terms of the posterior predictive $p$-value (ppp) and the posterior-averaged reduced chi-square $\langle  \chi^2_\nu \rangle_{\rm post}$. The ${\rm ppp}$ assumes values between 0 and 1, and we use the treshold ${\rm ppp}\geq 0.95$ to reject the model. The $\langle \chi^2_\nu \rangle_{\rm post}$ is compared to the 95 percent (upper) confidence limit associated to a number of degrees of freedom equal to the total number of data points fitted: when $\langle  \chi^2_\nu \rangle_{\rm post}$ is greater than this value, the model fails to describe the data. For the comparison between different models and assumptions on the bias parameters, we use the Deviance Information Criterion (DIC) computed from the MCMC simulations. For details on the choice of these diagnostics, we redirect the reader to \PaperI.

\section{Results}

\subsection{Maximal model}
\label{ssec:maximal_model}

We start with a test of the full model in which all seven bias and shot-noise parameters are free to vary. We compare the fit to the bispectrum monopole to the joint analysis of monopole and higher-order multipoles, assuming the full volume of the 298 Minerva simulations. The first goal is to identify the set of parameters that can effectively be determined by our data-set, and the relative importance of the different multipoles in setting their constraints.  We assume uniform priors on all parameters, with bounds specified in table~\ref{tab:priorsB}.

\begin{table}[t]
    \centering
    \begin{tabular}{ll|c}
        \hline
        \hline
        \multicolumn{2}{l}{Parameter} & Prior (uniform)\\
        \hline
        $b_1$ & & $[0.9,3.5]$\\
        $b_2$ & & $[-4,4]$\\
        $\bG$ & & $[-4,4]$\\
        $\alpha_1$ & &$[-1,2]$\\        
        $\alpha_2$ &  &$[-1,2]$\\
        $\alpha_3$ &  &$[-5,2]$\\
        $f$ & &$[0.1,1]$\\
        \hline
        \hline
    \end{tabular}
    \caption{Uniform prior intervals of the model parameters.}
    \label{tab:priorsB}
\end{table}

\begin{figure}[t]
    \centering
    \includegraphics[width=0.98\textwidth]{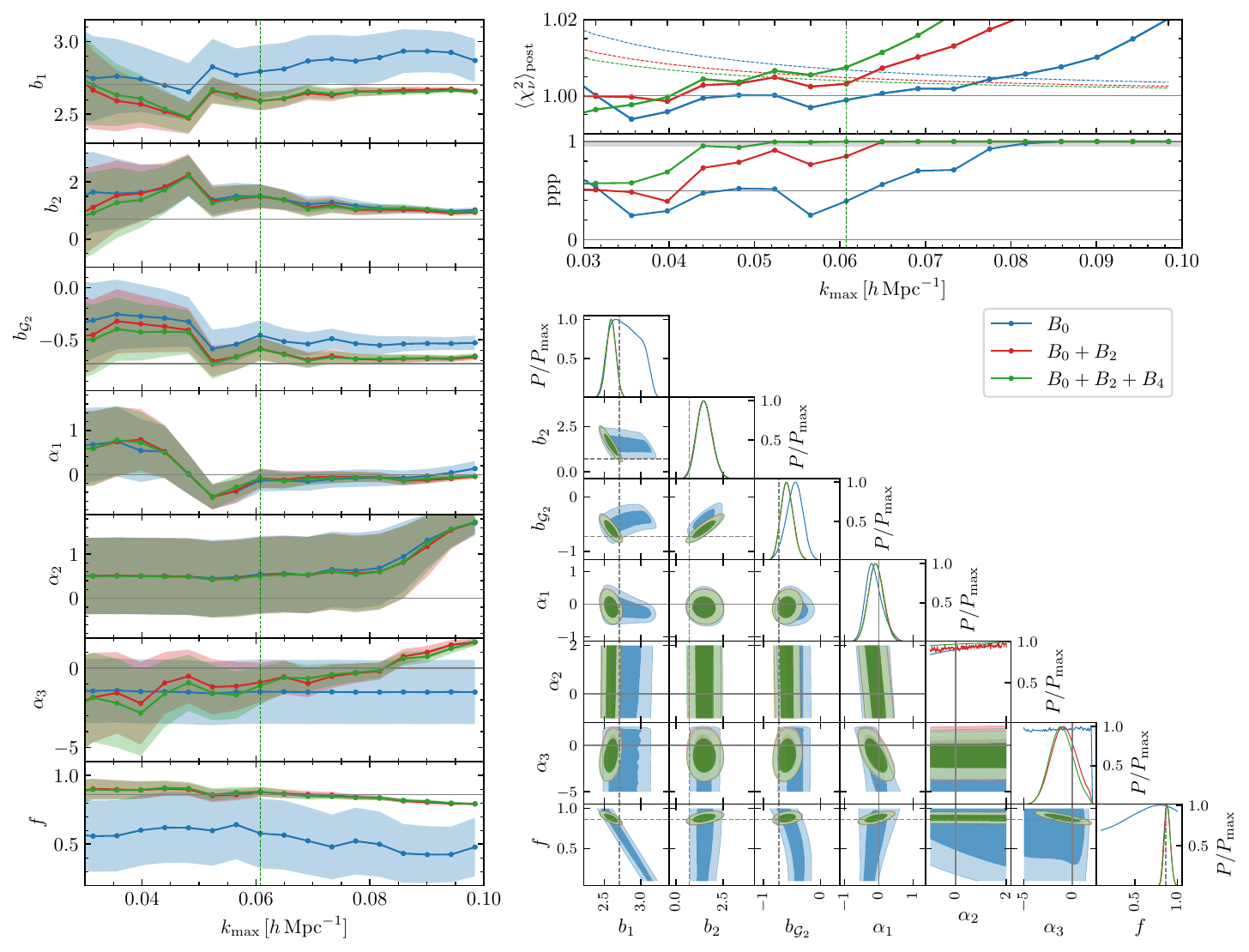}
    \caption{Results for the analysis of the whole 298 Minerva simulations data-set in terms of the full, seven-parameters model. Left panels: marginalised, 1-$\sigma$ posteriors for each parameter as a function of $\kmax$. Top-right panels: posterior-averaged, reduced chi-square, $\langle\chi_\nu^2\rangle$ and the posterior predictive $p$-value ({\rm ppp}) as a function of $\kmax$. The blue, red and green dashed lines in the $\langle\chi_\nu^2\rangle$ panel represent the 95\% confidence limits for the three combinations of multipoles considered. Bottom-right panel: two-dimensional, marginalised 1-$\sigma$ contours for $\kmax=0.06\kMpc$ case (corresponding to the vertical line in the other panels). In all panels, the $B_0$-only analysis (blue) is compared to the joint $B_0+B_2$ (red) and $B_0+B_2+B_4$ (green). All posteriors are compared with the results from the joint analysis of the real-space power spectrum and bispectrum derived in \PaperII{}, whose best-fit values are shown by the gray, dashed lines.}
    \label{fig:B0B2B4}
\end{figure}

The main results are shown in figure~\ref{fig:B0B2B4}. The left panels show the marginalised, 68.3\% credibility regions for the model parameters as a function of the maximum wavenumber included in the triangle selection. Two-dimensional marginalised contours are shown in the bottom-right panel for the $\kmax=0.06\kMpc$ case (indicated as the vertical, dotted line in the other two sub-panels). Finally, the top-right panel shows the posterior-averaged, reduced chi-square, $\langle\chi_\nu^2\rangle$ and the posterior predictive $p$-value.   

The tree-level model described in section~\ref{ssec:model} provides a good fit to the data up to a $\kmax\simeq 0.08\kMpc$ for the monopole, while the reach is restricted to 0.06 and 0.045$\kMpc$, respectively, when $B_2$ and $B_4$ are also considered. 

For $\kmax=0.06\kMpc$, the combination $B_0+B_2$ properly recovers the best-fit values of the bias parameters obtained from the joint analysis of the real-space power spectrum and bispectrum along with the fiducial value of the linear growth rate $f$, which are shown by the gray, dashed lines in the bottom-right panel.

The addition of the quadrupole $B_2$ greatly reduces the large degeneracy between the linear bias $b_1$ and the growth rate $f$, as well as with the quadratic bias parameters, characterising the monopole-only constraints. This qualitatively confirms the expectation that the bispectrum monopole does not fully capture the information potentially present in the anisotropic, redshift-space bispectrum \cite{YankelevichPorciani2019, GualdiEtal2020}. On the other hand, including as well the hexadecapole leads to no significant improvement for any parameter and we will drop it for all the tests that will follow. 

\subsection{Shot-noise}

It is clear that, despite the large total simulation volume, the data set is not able to provide meaningful constraints on all shot-noise parameters. This was true as well, for the same halo catalogs, in real space, even including power spectrum information (see \PaperII{}). 

In this section we compare different options to reduce the shot-noise parameters to a single one. In addition to the maximal model, characterised by seven parameters in total, we will consider the following models 
\begin{itemize}
    \item $\alpha_2=0$ (6 parameters); this is justified by the posteriors obtained for the maximal model, which show that this parameter is simply not constrained by the data and can therefore be set to zero without affecting the overall fit;      
    \item $\alpha_3=-1$ (6 parameters); this corresponds to setting $\epsilon_\eta=0$, as expected under the assumption of no velocity bias and no selection effects \citep{DesjacquesJeongSchmidt2018B}.
\end{itemize}
Another option is to set $\alpha_1=\alpha_3$. This is implicit in the Poisson prediction for the shot-noise of a generic distribution in redshift space, where {\em both} corrections vanish. This prediction is also the outcome of a count-in-cell estimate of the shot-noise contributions to the bispectrum \cite{Peebles1980, MatarreseVerdeHeavens1997} and corresponds to the standard shot-noise correction often implemented in bispectrum estimators \cite{Scoccimarro2015, SugiyamaEtal2019} and implicitly assumed in some data analysis \cite{GilMarinEtal2015, GilMarinEtal2017}. We will therefore consider the two, additional 5-parameters models (both assuming $\alpha_2=0$):
\begin{itemize}
    \item $\alpha_3=\alpha_1$ {\em and} $\alpha_2=0$ (5 parameters); 
    \item $\alpha_3=-1$ {\em and} $\alpha_2=0$ (5 parameters). 
\end{itemize}

\begin{figure}[t!]
    \centering
    \includegraphics[width=0.91\textwidth]{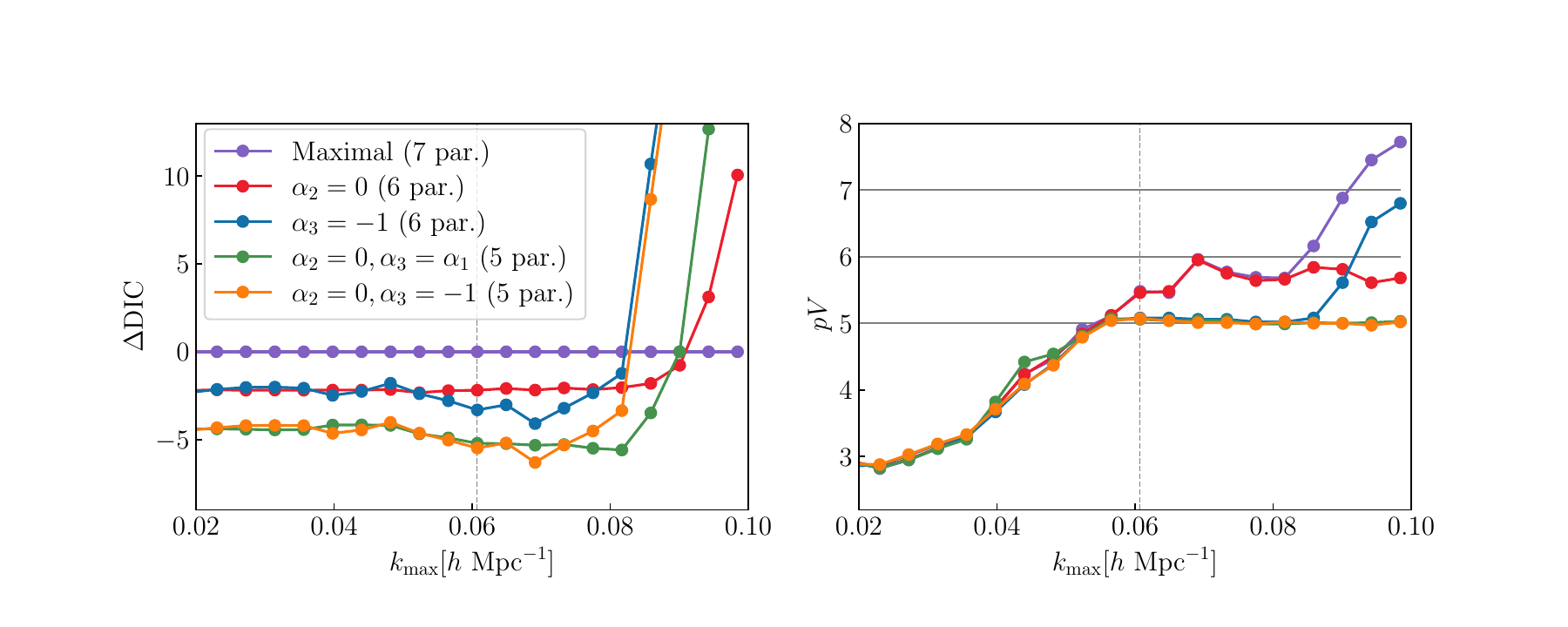}
    \includegraphics[width=0.75\textwidth]{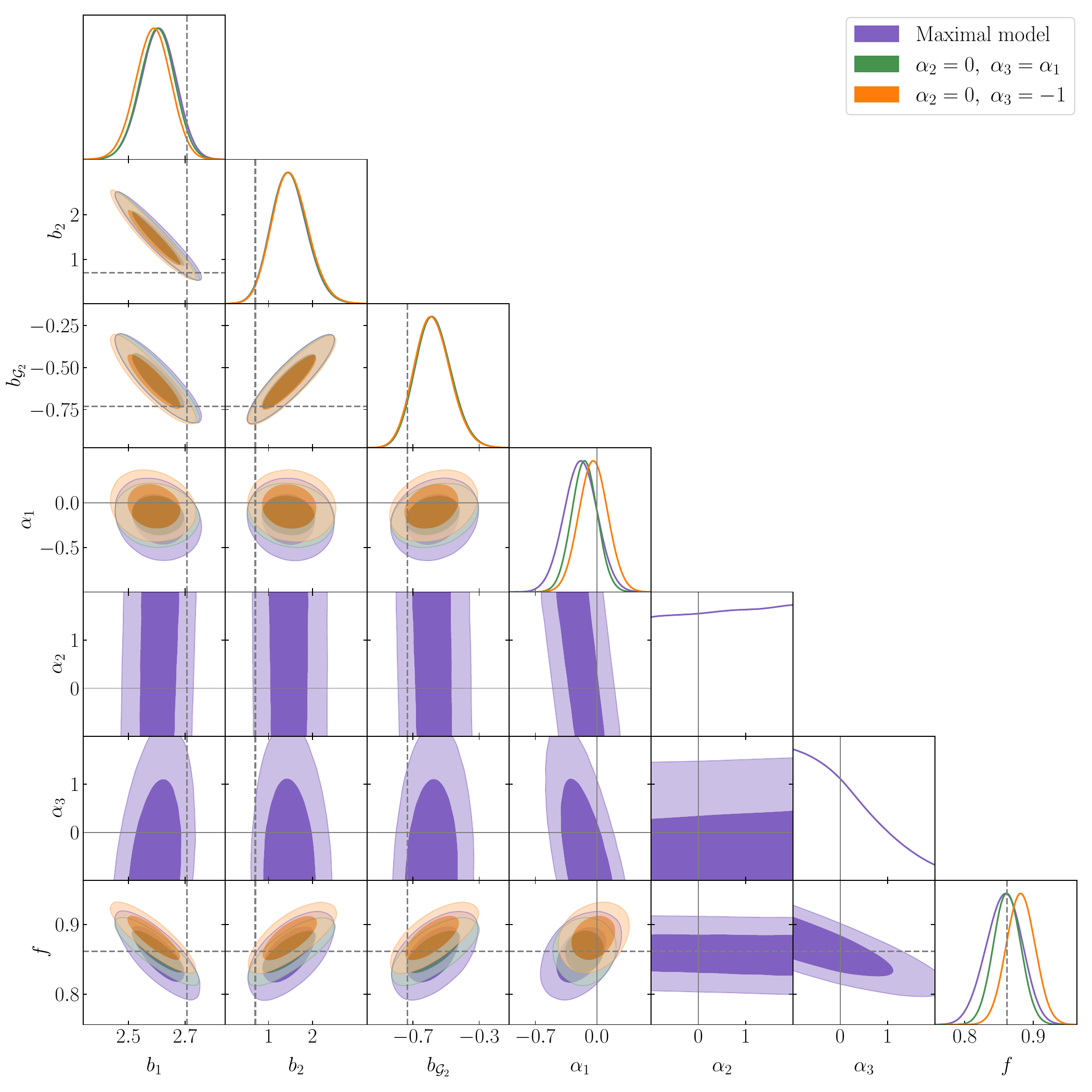}
    \caption{Top-left panel: difference in the DIC of the various shot-noise models with respect to the maximal one with seven free parameters, as a function of the largest wavenumber $\kmax$ for the analysis of the monopole plus quadrupole. Top-right panel: effective numbers of parameters as a function of $\kmax$ for the same shot-noise models and the same dataset, $B_0+B_2$. Bottom panel: contour plots for $B_0+B_2$ at $\kmax=0.06\kMpc$, showing the 68.3 and 95.4\% credible intervals on the bias and shot-noise parameters for the maximal model compared to the two 5-parameter models defined by setting $\alpha_3=\alpha_1$ and $\alpha_3=-1$ (both with $\alpha_2=0$). Dashed lines represent the true value of growth rate $f$ and the best-fit values for the bias parameters obtained in real space in \PaperII.}
    \label{fig:shot-noise}
\end{figure}

The top left panel in figure~\ref{fig:shot-noise} shows a general comparison between all the models described in the bullet points above in terms of the difference in their DIC w.r.t. the maximal model with seven parameters, as a function of the maximum wavenumber $\kmax$, again for the monopole and quadrupole analysis. Differences larger than 5 are usually considered relevant. The top right panel of the same figure shows instead the effective number of parameters we are able to constrain from the data also as a function of $\kmax$. For a value of $\kmax<0.05 \kMpc$, we do not have enough information to determine even 5 parameters and the $\Delta$DIC simply favours the simplest models. These are still favoured up to $\kmax\sim 0.08\kMpc$, where the additional degrees of freedom of more complex models are probably accounting for missing nonlinear corrections. The test does not clearly indicate a preference for the 5-parameter model with $\alpha_3=\alpha_1$ over the one with $\alpha_3=-1$, except for $\kmax>0.08\kMpc$, where we know that none of the models provides a good overall fit anymore. A comparison of the 2D marginalised posteriors from the monopole and quadrupole analysis at $\kmax=0.06\kMpc$ is shown for the two 5-parameters models and the maximal one in the bottom panel of figure~\ref{fig:shot-noise}. Both models improve the constraints on the growth rate, with minimal differences on the posteriors for the other parameters. The $\alpha_3=-1$ case provides a slightly better agreement with the fiducial value of $f$ and the real-space estimate of $b_1$. 

We will assume the $\alpha_3=-1$ (i.e. $\epsilon_\eta=0$) and $\alpha_2=0$ case as our default model in all following tests. This implies the expression for the shot-noise contribution  
\begin{align}
\label{eq:B_stochastic_final}
B_s^{\rm (stoch)}(\kv_1, \kv_2, \hat{n})  =  \frac{1+\alpha_1}{\bar{n}}\,b_1\,Z_1(\kv_1)\,P_L(k_1)+2~{\rm perm.}
+\frac{1}{\bar{n}^2}\,,
\end{align}
only depending on the parameter $\alpha_1$. We assume that this model provides an accurate description of the stochastic contribution to the bispectrum, consistent with the large-scale expectation, in a relatively restricted range ($k \lesssim 0.08\kMpc$) where the Poisson limit ($\alpha_1=\alpha_3=0$) does not apply.

\subsection{Bias relations}

The parameter space can be further reduced by introducing relations among the bias parameters. In \PaperI{} and \PaperII{} we considered a few of them, either theoretically motivated or extracted from numerical simulations \cite{ShethChanScoccimarro2013, LazeyrasEtal2016, LazeyrasSchmidt2018, EggemeierEtal2020, EggemeierEtal2021}.  Of those, we select the two that provide the best improvement to the fit of the power spectrum and bispectrum in real space and test them again here in redshift space. The first is the fitting function for $b_2(b_1,\bG)$ obtained in \cite{LazeyrasEtal2016} from separate universe simulations. The second is the fit to the excursion set prediction for the tidal bias parameter $\bG(b_1)$ proposed in \cite{EggemeierEtal2020, EggemeierEtal2021}.  For convenience we reproduce these two relations here:
\begin{align}
    b_2(b_1,\bG) &= 0.412 - 2.142\,b_1 + 0.929\,b_1^2 + 0.008\,b_1^3 + \frac{4}{3}\bG\,, \\
    \bG(b_1) &= 0.524 - 0.547\,b_1 + 0.046\,b_1^2\,.
\end{align}

\begin{figure}[th]
    \centering
    \includegraphics[width=0.91\textwidth]{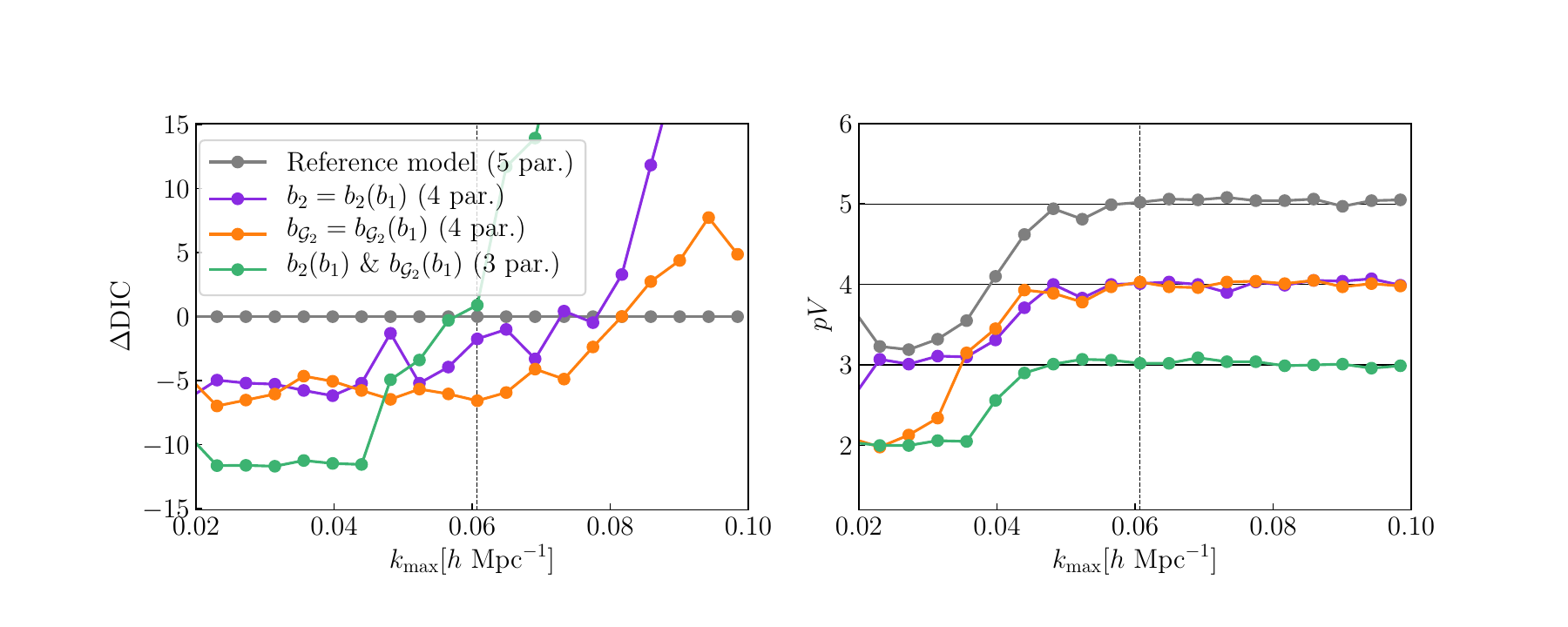}    
    \includegraphics[width=0.75\textwidth]{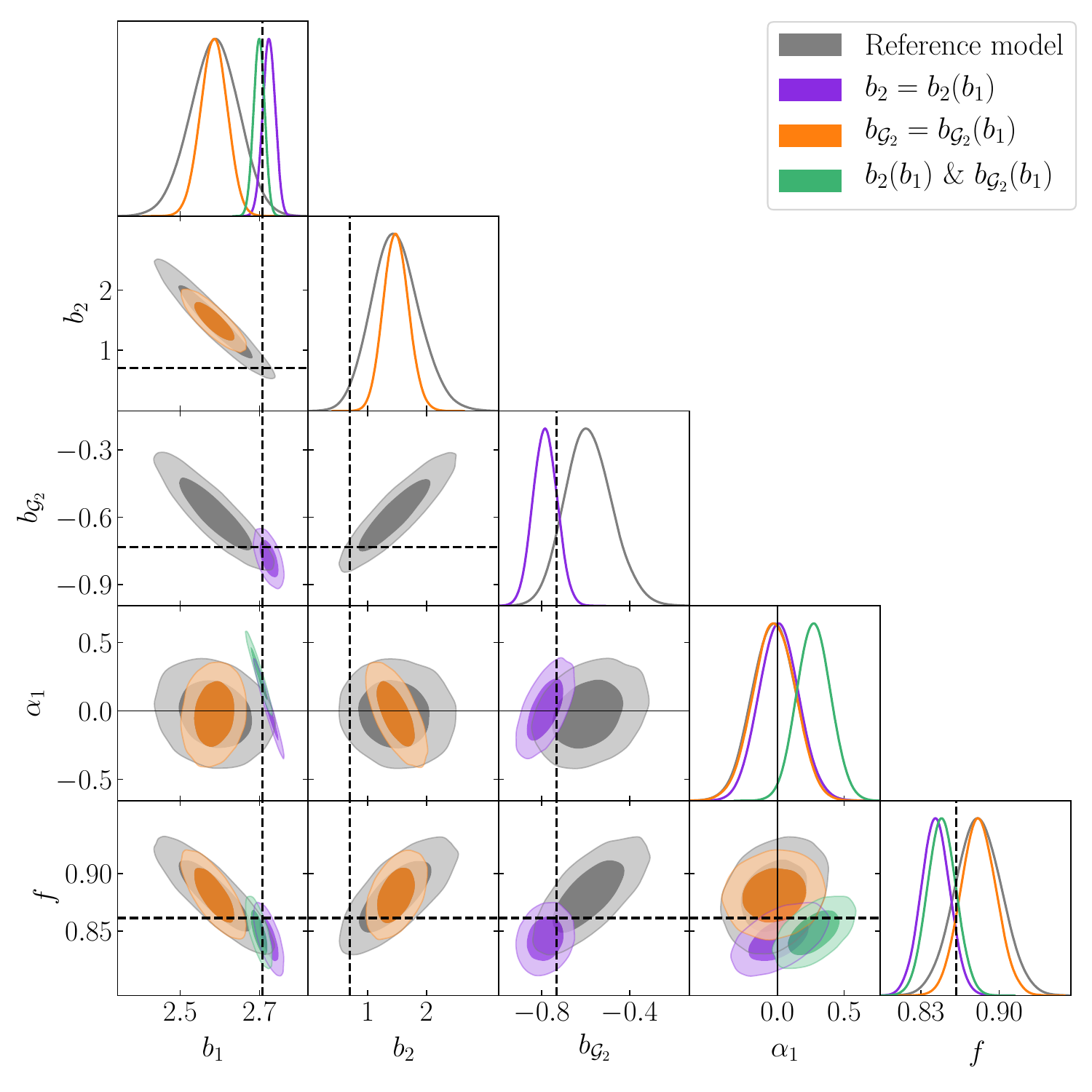}    
    \caption{Top-left panel: difference in the DIC of the two models adopting the bias relations $b_2(b_1,\bG)$ and $\bG(b_1)$  with respect to the reference, 5-parameter model, as a function of the largest wavenumber $\kmax$ for the analysis of the monopole plus quadrupole. Top-right panel: effective numbers of parameters as a function of $\kmax$ for the same models and datasets, $B_0+B_2$. Bottom panel: contour plots for $B_0+B_2$ at $\kmax=0.06\kMpc$, showing the 68.3\% and 95.4\% credible intervals on the bias and shot-noise parameters for the three models. Dashed lines represent the true value of growth rate $f$ and the best-fit values for the bias parameters obtained in real space in \PaperII.}
    \label{fig:bias_relations}
\end{figure}

In figure~\ref{fig:bias_relations} we compare three cases with our reference 5-parameter model: applying each of the bias relations $b_2(b_1,\bG)$ and $\bG(b_1)$ individually, as well as the two of them combined. In the top-left panel we show their difference in the DIC with respect to the reference model, as a function of the largest wavenumber $\kmax$ for the analysis of the monopole plus quadrupole. The top-right panel shows instead the effective numbers of parameters, again as a function of $\kmax$ and for $B_0+B_2$. The DIC shows a marginal preference for the $\bG(b_1)$ relation, for values of $\kmax$ close to $0.06\kMpc$, whereas the combination of the two relations quickly becomes disfavoured beyond $\kmax = 0.05\kMpc$.

In the bottom panel of figure~\ref{fig:bias_relations} we show the contour plots from the analysis at $\kmax=0.06\kMpc$, showing the 68.3\% and 95.4\% credible intervals on the bias and shot-noise parameters for the three models. All three cases lead to tighter marginalised posterior contours, however, we notice how the application of the tidal bias relation leads to constraints on $b_1$ and $f$ that are systematically offset from the fiducial values, while the other two cases involving the $b_2(b_1)$ relation significantly reduces any potential tension. We caution that this outcome might in fact be fortuitous since the $b_2(b_1)$ relation crosses the $b_1$ - $b_2$ contour of the reference model close to the fiducial value of $b_1$ recovered from the real-space analysis of \PaperII{}. We should stress, in any case, that these systematic differences are only evident due to the very large cumulative simulation volume: we leave for future work an assessment for a volume that can be achieved in real surveys \cite{MorettiEtalInPrep} .

\subsection{Scale cuts}

In section~\ref{ssec:maximal_model} we have shown that the range of validity of the tree-level model is more restricted for higher-order multipoles with respect to the monopole, i.e. the model starts failing at larger scales. This suggests the possibility to adopt different values of $\kmax$ for the different multipoles, pushing the bispectrum monopole to smaller scales. 

\begin{figure}[t!]
    \centering
    \includegraphics[width=0.75\textwidth]{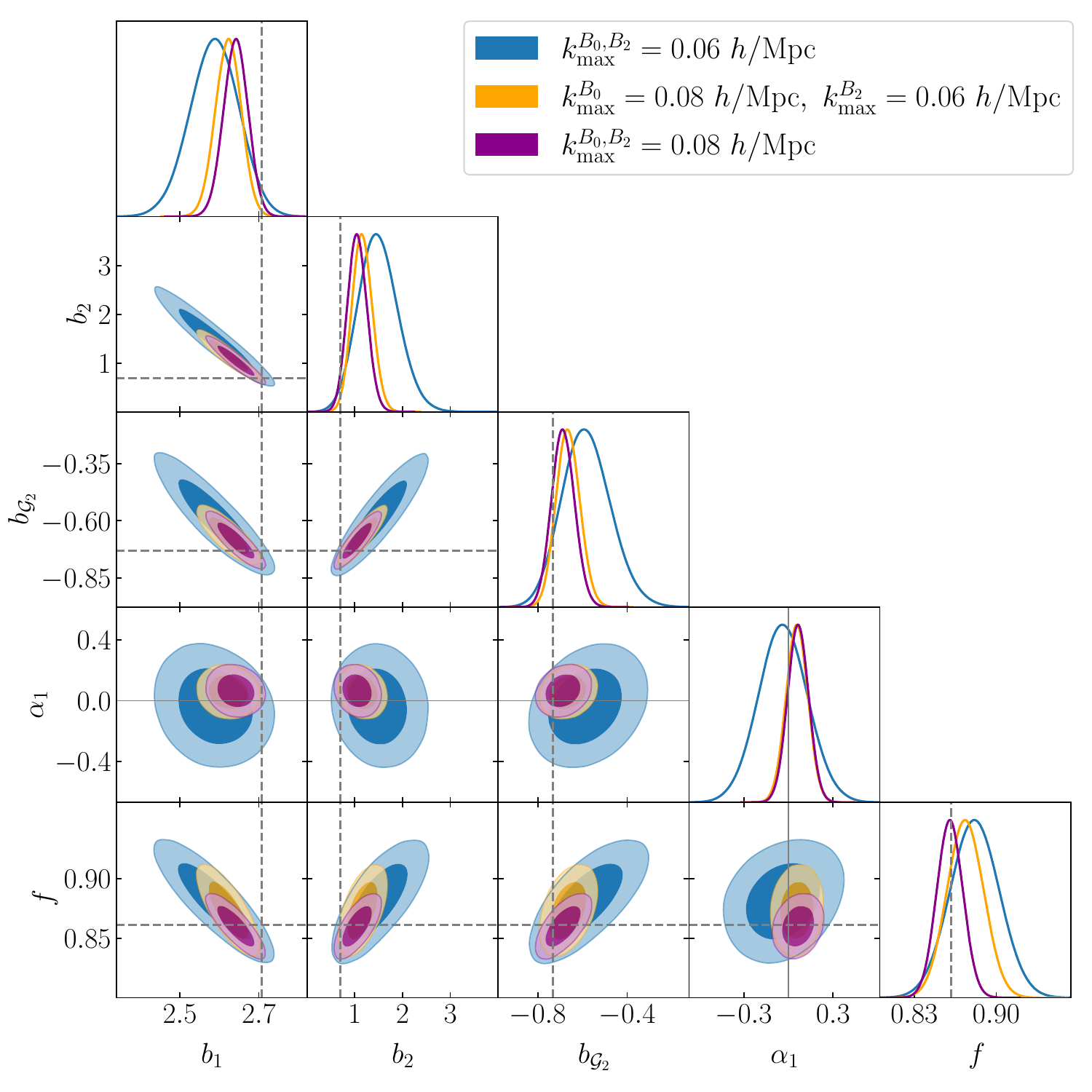}
    \caption{Contour plots for the $B_0+B_2$ analysis assuming $\kmax^{B_0}=\kmax^{B_2}=0.06\kMpc$ (blue) compared to the cases with $\kmax^{B_0}=0.08\kMpc$ and $\kmax^{B_2}=0.06\kMpc$ (yellow) and $\kmax^{B_0}=\kmax^{B_2}=0.08\kMpc$ (magenta). The gray, dashed mark the best-fit values from the real-space, joint analysis of power spectrum and bispectrum of \PaperII{}.}
    \label{fig:scale_cuts}
\end{figure}

We illustrate this point in figure~\ref{fig:scale_cuts} where we compare the contour plots for the $B_0+B_2$ analysis under the following assumptions:
\begin{itemize}
    \item $\kmax^{B_0}=\kmax^{B_2}=0.06\kMpc$ (blue);
    \item $\kmax^{B_0}=0.08\kMpc$ and $\kmax^{B_2}=0.06\kMpc$ (yellow);
    \item $\kmax^{B_0}=\kmax^{B_2}=0.08\kMpc$ (magenta).
\end{itemize}
In all cases we adopt the reference, 5-parameter model. 

We find that extending $\kmax^{B_0}$ to $0.08\kMpc$ can significantly reduce the errors on the bias parameters in particular. However, for the $\kmax^{B_0}=\kmax^{B_2}=0.08\kMpc$ case, where the $\chi^2$ for the fit is already above the 95\% C.L., we notice that the $f$-$b_1$ contour already shows a discrepancy with the expected values at more than 95.4\% credible regions. We will adopt the scale cuts defined by $\kmax^{B_0}=0.08\kMpc$ and $\kmax^{B_2}=0.06\kMpc$ as our reference choice for most of the tests in the following sections.

\subsection{Covariance approximations}

In section~\ref{ssec:covariance} we directly compared the Gaussian prediction for the bispectrum covariance with the numerical estimate from the \pin{} mocks, finding a remarkable agreement both in the variance $C_{\ell_1\ell_2}(t_i,t_i)$ as in the correlation coefficients  $r_{\ell_1\ell_2}(t_i,t_i)$. 

\begin{figure}[t!]
    \centering
    \includegraphics[width=0.75\textwidth]{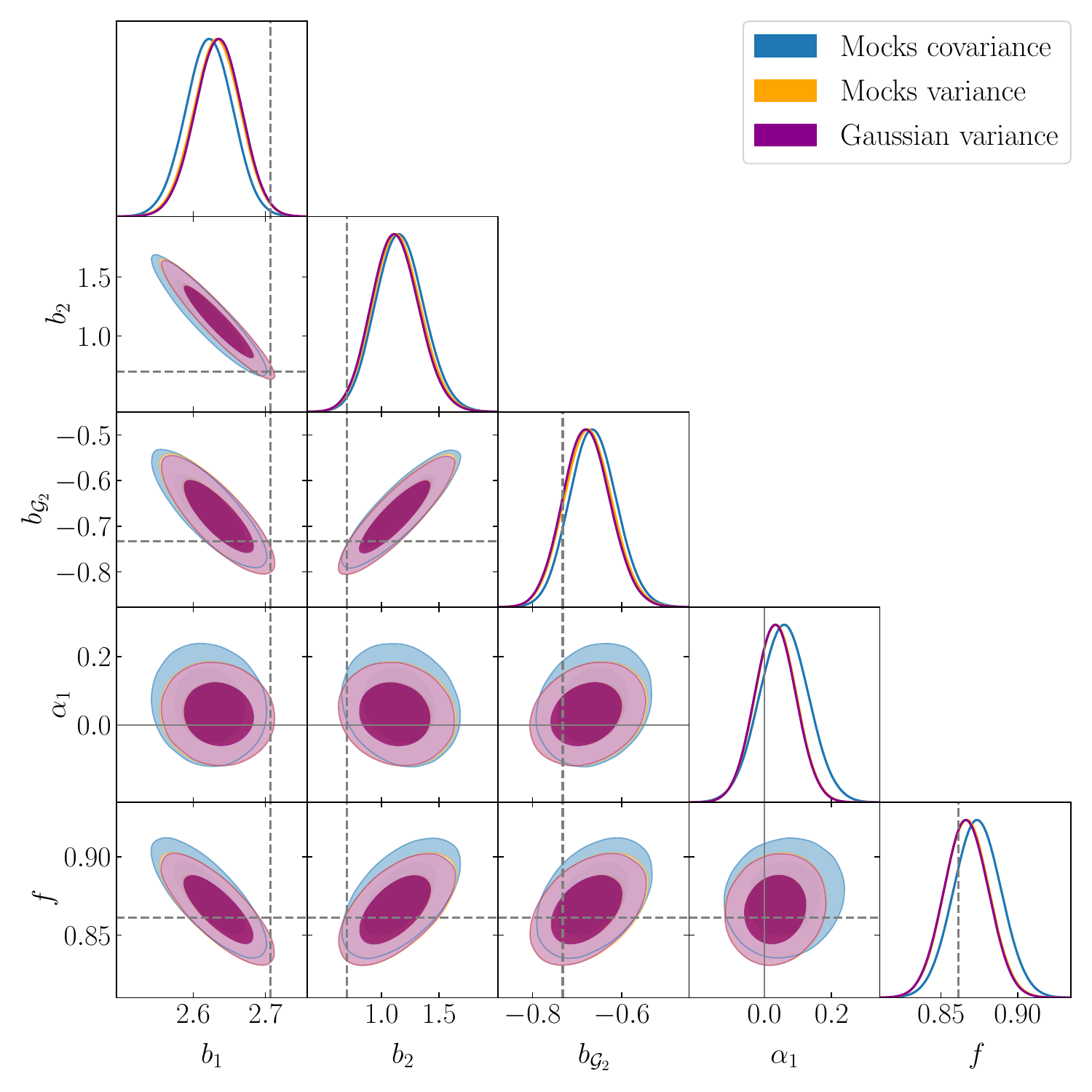}    
    \caption{Contour levels of the marginalised posterior distribution for the $B_0+B_2$ reference analysis under different assumptions for the covariance. The case of the full covariance, $C_{\ell_1\ell_2}(t_i,t_j)$, estimated from the mocks (blue contours) is compared to the variance $C_{\ell_1\ell_2}(t_i,t_i)$ estimated from the mocks (orange) and from the Gaussian theory prediction (magenta). The last two cases overlap almost exactly. The analysis assumes $\kmax^{B_0}=0.08\kMpc$ and $\kmax^{B_2}=0.06\kMpc$.}
    \label{fig:cov_vs_variance}
\end{figure}

Here we check if any residual difference could affect the parameters determination. The comparison is shown in figure~\ref{fig:cov_vs_variance} in terms of the contour plots for the reference analysis of $B_0$ and $B_2$ assuming, respectively, $\kmax^{B_0}=0.08\kMpc$ and $\kmax^{B_2}=0.06\kMpc$. The full numerical covariance, $\hat{C}_{\ell_1,\ell_2}(t_i,t_j)$ (blue contours) is compared to the numerical variance $\hat{C}_{\ell_1,\ell_2}(t_i,t_i)$ (orange) and to the Gaussian prediction (magenta). In the last two cases, all elements with $t_i\ne t_j$ are set to zero. We notice that the inclusion of such elements is responsible for constraints only slightly larger, while the Gaussian prediction reproduces the results from the numerical variance case almost exactly, with no appreciable differences in the 1D or 2D marginalised posteriors. This is perhaps not too surprising given that the analysis is restricted to relatively large-scales.

\subsection{Effective binning of the theoretical model }
\label{app:EffectiveBinning}

All of our results assumed an evaluation of the theory predictions implementing the exact scheme of eq.~(\ref{e:exact_binning}). Since this approach can be numerically quite demanding, particularly in likelihood evaluations extended to several cosmological parameters, it is worth exploring the systematic errors induced on the parameter posteriors by the more efficient choice of a single bispectrum evaluation on the effective wavenumbers, after the analytical integration over the angles described in Appendix~\ref{app:A}.

In this case, the theoretical prediction is given by
\be
\label{e:effective_binning}
B_{\ell}^{\rm eff}(k_1, k_2, k_3) \equiv B_{\ell}(k_{\rm eff,l}, k_{\rm eff,m}, k_{\rm eff,s}),
\ee
where the definition of the effective triplet, in general not unique, is based on ``sorted'' $\left\{q_1,q_2,q_3\right\}$ triplets as (see \PaperI{})\footnote{See also \cite{EggemeierEtal2021} and \cite{IvanovEtal2021A} for alternative proposals.}.   
\begin{align}
k_{\rm eff, l}(k_1, k_2, k_3) & =\frac{1}{N_{B}}\sum_{\textbf{q}_1\in k_1}\sum_{\textbf{q}_2\in k_2}\sum_{\textbf{q}_3\in k_3} \delta_K(\textbf{q}_{123})\max(q_1,q_2,q_3)\,,
\nn \\
k_{\rm eff, m}(k_1, k_2, k_3) & =\frac{1}{N_{B}}\sum_{\textbf{q}_1\in k_1}\sum_{\textbf{q}_2\in k_2}\sum_{\textbf{q}_3\in k_3} \delta_K(\textbf{q}_{123}){\rm med}(q_1,q_2,q_3)\,,
\nn \\
k_{\rm eff, s}(k_1, k_2, k_3) & =\frac{1}{N_{B}}\sum_{\textbf{q}_1\in k_1}\sum_{\textbf{q}_2\in k_2}\sum_{\textbf{q}_3\in k_3} \delta_K(\textbf{q}_{123})\min(q_1,q_2,q_3)\,.
\end{align}

\begin{figure}[t!]
    \centering
    \includegraphics[width=0.75\textwidth]{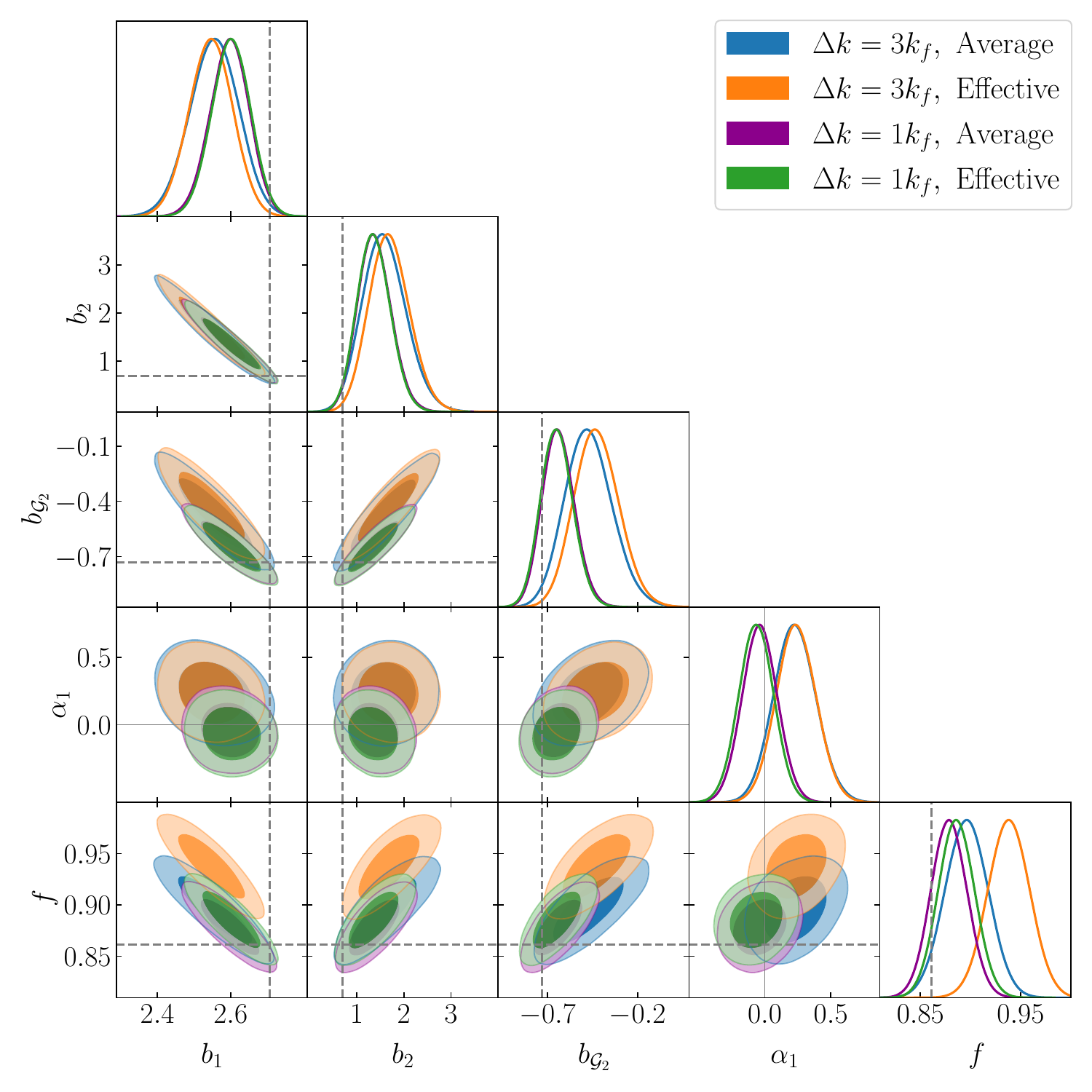}    
    \caption{Contour levels of the marginalised posterior distribution for the $B_0+B_2$ reference analysis for $\kmax^{B_0}=\kmax^{B_2}=0.06\kMpc$, under different assumptions for the binning scheme and evaluation of the theoretical prediction: $\Delta k = k_f$ with exact binning (magenta), $\Delta k = k_f$ with effective prediction (green), $\Delta k = 3~k_f$ with exact binning (blue) and $\Delta k = 3~k_f$ with effective prediction (orange).}
    \label{fig:binning_vs_effective}
\end{figure}

For the choice of the bin size $\Delta k = k_f$ adopted in our result, the difference between the two approaches, estimated in terms of the posteriors on the bias and shot-noise parameters, is completely negligible. For a larger size of the bin $\Delta k$, useful to reduce the overall size of the data vector, however, we can find some systematic effect on parameters determination.  This is shown in fig.~\ref{fig:binning_vs_effective}, where we plot the 2D marginalised posteriors for different choices of the binning scheme and evaluations of the theoretical prediction. In particular, there is a significant shift in the 1D marginalised posterior for $f$ (shown in orange in fig.~\ref{fig:binning_vs_effective}) for the case $\Delta k = 3~k_f$ when we compute the theoretical prediction at the effective wavenumbers. We notice as well how the larger bin size leads to weaker constraints, even in the exact binning case, due to the reduced shape-dependence of the bispectrum measurements.

\section{Conclusions}

This work presents a test of the tree-level prediction in Perturbation Theory for the halo bispectrum in redshift space with particular attention to its anisotropic signal as described by higher-order multipoles such as the quadrupole and hexadecapole. It extends previous results in real space (\PaperI{} and \PaperII{}), taking advantage of a very large set of 298 N-body simulations corresponding to a cumulative volume of about 1,000$\cGpc$ and an even larger set of \pin{} mocks.  The latter provides a robust estimate of the covariance properties for the full data vector given by the three bispectrum multipoles.
We explore different assumptions on the observables and related covariance models and assess them in terms of constraints on bias parameters and the linear growth rate.  

We summarise below our main findings.
\begin{itemize}
    \item The \pin{} mocks provide a very good description of the variance estimated from the full numerical simulations with a residual scatter below the 10\% level and no apparent shape-dependence (Fig.~\ref{fig:variance}), for all bispectrum multipoles, extending previous assessments in real-space (\PaperI{}, \cite{ColavincenzoEtal2019}).

    \item The comparison of the posterior distributions based on the monopole alone with the joint analyses of $B_0+B_2$ and $B_0+B_2+B_4$ (Fig.~\ref{fig:B0B2B4}), using the full covariance from the mocks, indicates that the addition of the quadrupole alone greatly improves the determination of bias parameters and, perhaps not surprisingly, allows to properly constrain the growth rate $f$; the further addition of the hexadecapole, instead, leads to no appreaciable improvement.

    \item For our full simulation volume, the tree-level model provides a good fit to the bispectrum monopole up to $\kmax=0.08\kMpc$, while the inclusion of the quadrupole and the hexadecapole reduce significantly this range to $0.06$ and $0.045\kMpc$, respectively. Indeed, an optimal configuration for the joint $B_0+B_2$ analysis should assume distinct values for $\kmax$. We tested that better constraints on the model parameters are obtained assuming  $\kmax^{B_0}=0.08\kMpc$ and $\kmax^{B_2}=0.06\kMpc$ with respect to the case where a single, lower value of $\kmax^{B_0}=\kmax^{B_2}=0.06\kMpc$ is adopted to keep within the model validity range (Fig.~\ref{fig:scale_cuts}).   
    \item In general our data, despite the large volume, cannot fully determine all shot-noise parameters describing different departures from the Poisson expectation. It appears, however, that the stochastic velocity contribution $\epsilon_\eta \eta$ to the halo density, that one can expect when selection effects are present \cite{DesjacquesJeongSchmidt2018}, is indeed disfavoured in our ideal set-up, at least at the large scales we are exploring (Fig.~\ref{fig:shot-noise}).

    \item Both the fitting function for the quadratic local bias parameter $b_2(b_1,\bG)$ of \cite{LazeyrasEtal2016} as the relation for the tidal bias parameter $\bG(b_1)$ proposed in \cite{EggemeierEtal2020, EggemeierEtal2021} (and their combination)  appear to significantly tighten the posteriors on $b_1$ and $f$ (Fig.~\ref{fig:bias_relations}); the first, however, appears to introduce some bias in the determination of $b_1$, although relevant only because of the large cumulative volume of our simulations set.

    \item For our ideal measurements in a simulation box with periodic boundary conditions, the Gaussian model for the covariance of the bispectrum multipoles provides a very good approximation to the numerical estimate. A small underestimate is noticeable (and expected \cite{Barreira2019, BiagettiEtal2021A}) for the squeezed configurations of the bispectrum monopole (Fig.~\ref{fig:variance_theory}). On the other hand a quite remarkable agreement is obtained in the comparison with the cross-correlation coefficients (Fig.~\ref{fig:cov}). In terms of the posteriors of the bias parameters and $f$ we find no noticeable difference between the Gaussian theory variance and the numerical variance, while a very small difference is present when compared to the full numerical covariance (Fig.~\ref{fig:cov_vs_variance}). 
    
    \item All our main results assume an exact binning of the theoretical prediction. When a single evaluation on a triplet of effective wavenumbers is assumed we notice a negligible difference only if the bin size is small, equal to the box fundamental frequency (Fig.~\ref{fig:binning_vs_effective}). For a larger bin size, in addition to somehow larger posteriors, this approximation can lead to much more significant shifts in the posteriors, resulting in evident systematic differences particularly on the growth rate $f$. 
\end{itemize}

As mentioned in the introduction, not many works explored the modelling and the information content of the anisotropic bispectrum, in particular in terms of constraints on bias parameters and the growth rate $f$ using Bayesian analyses. Our results qualitatively confirm earlier Fisher-matrix forecasts \cite{SongTaruyaOka2015, GagraniSamushia2017, YankelevichPorciani2019, AgarwalEtal2021, ByunKrause2022} in remarking the importance of going beyond the analysis of the bispectrum monopole. The same can be said for \cite{GualdiVerde2020} and \cite{GualdiGilMarinVerde2021}, both based instead on a full likelihood analysis and therefore closer, in principle, to our work. For these last two references, however, many differences in methodology do not allow a rigorous, quantitative comparison with our results, in addition to the fact, of course, that we do not include power spectrum measurements in our data-vector. We will present a joint analysis of the Minerva-set power spectrum and bispectrum in redshift space elsewhere \cite{MorettiEtalInPrep}. For the time being we limit ourselves to observe that the inclusion of full anisotropic bispectrum information will likely be an important step toward a complete exploitation of cosmological information in spectroscopic galaxy surveys.   

\acknowledgments

We are always grateful to Claudio Dalla Vecchia and Ariel Sanchez for running and making available the Minerva simulations, performed on the Hydra and Euclid clusters at the Max Planck Computing and Data Facility (MPCDF) in Garching. We are grateful to Guido d'Amico and Vincent Desjacques for useful discussions. The Pinocchio mocks were run on the GALILEO cluster at CINECA, thanks to an agreement with the University of Trieste. K.P., E.S. and P.M. are partially supported by the INFN INDARK PD51 grant and acknowledge support from PRIN MIUR 2015 Cosmology and Fundamental Physics: illuminating the Dark Universe with Euclid. C.M. acknowledges support from a UK Research and Innovation Future Leaders Fellowship (MR/S016066/1).  For the purpose of open access, the author has applied a Creative Commons Attribution (CC-BY) licence to any Author Accepted Manuscript version arising from this submission.

\appendix

\section{Analytical evaluation of the bispectrum multipoles}
\label{app:A}

The orientation of the $\left\{\kv_1, \kv_2, \kv_3\right\}$ triangle w.r.t. the LOS in the model for the redshift-space bispectrum $B_s(\kv_1, \kv_2, \hat{n})$, eq.~(\ref{eq:Bmodel}), is expressed entirely in terms of products of powers of the cosines $\mu_i\equiv \hat{k}_i\cdot\hat{n}$ that can be factorised in each contribution.

The integrals defining the bispectrum multipoles in eq.~(\ref{e:Bl}) are therefore limited to angle-averages of such combinations, with the additional factors from the Legendre polynomials corresponding to additional powers of $\mu_1$. We denote these integrals as  
\be
\label{e:Iabc}
\mathcal{I}_{\alpha \beta \gamma} = \frac1{4\pi}\int_{-1}^{+1} d\mu_1 \int_{0}^{2\pi} d\xi\, \mu_1^{\alpha}\,\mu_2^{\beta}(\mu_1, \xi)\,\mu_3^{\gamma}(\mu_1, \xi)\,,
\ee
where
\be
\mu_2(\mu_1,\xi) = \mu_1\mu_{12}-\sqrt{1-\mu_1^2}\sqrt{1-\mu^2_{12}}\cos\xi
\ee
and 
\be
\mu_3(\mu_1,\xi) = - \frac{k_1\mu_1 + k_2\mu_2(\mu_1,\xi)}{k_3}\,,
\ee
having introduced $\mu_{12} \equiv \hat{k}_1\cdot\hat{k_2}$. Since the angle integration of eq.~(\ref{e:Iabc}) is to be intended as a generic integration over all orientations, it is easy to see that it should satisfy the following property 
\be
\I_{\alpha,\beta,\gamma}(k_1,k_2,k_3)=\I_{\sigma(\alpha,\beta,\gamma)}[\sigma(k_1,k_2,k_3)]\,,
\ee
where $\sigma(...)$ represents a generic permutation applied, {\em at the same time}, to its indices {\em and} arguments\footnote{Notice that in general is not true that $\I_{\alpha,\beta,\gamma}(k_1,k_2,k_3)=\I_{\sigma(\alpha,\beta,\gamma)}(k_1,k_2,k_3)$, so if we write, e.g.,  $\I_{200}+\I_{020}$ we implicitly mean $\I_{200}(k_1,k_2,k_3)+\I_{020}(k_1,k_2,k_3)$ corresponding to the sum of two different quantities.} We write, for illustration purposes the first few as
\begin{align}
\I_{\alpha 00} & = 
\frac{1}{1+\alpha} ~ {\rm for}~ \alpha~{\rm even~ (vanishing~ otherwise)}\,,\\
\I_{\alpha 01} & = -
\frac{k_1^2+k_3^2-k_2^2}{2(2+\alpha)\,k_1\,k_3} ~ {\rm for}~ \alpha~{\rm odd~ (vanishing~ otherwise)}\,,\\
\I_{\alpha 11} & = \frac{(2+\alpha)\,k_1^4-\alpha\,(k_2^2-k_3^2)^2-2\,k_1^2\,(k_2^2+k_3^2)}{4(1+\alpha)(3+\alpha)k_1^2\,k_2\,k_3} ~ {\rm for}~ \alpha~{\rm even~ (vanishing~ otherwise)}\,,\\
\I_{\alpha 02} & = \frac{4\,k_1^2\,k_3^2+\alpha\,(k_1^2+k_3^2-k_2^2)^2}{4(1+\alpha)(3+\alpha)k_1^2\,k_3^2} ~ {\rm for}~ \alpha~{\rm even~ (vanishing~ otherwise)}\,.
\end{align}

We can group all contributions to the bispectrum multipoles according to the source of quadratic nonlinearity, that is 
\be
B_\ell(k_1,k_2,k_3) =   B_{\ell}^{(F_2)}+B_{\ell}^{(b_2)} + B_{\ell}^{(S_2)} + B_{\ell}^{(G_2)} + B_{\ell}^{({\rm mixed})}\,.
\ee
These contributions can be expressed in terms of the $\I_{\alpha\beta\gamma}$
\begin{align}
\label{e:B0F2}
B_{0}^{(F_2)} & = 
2\,b_1^3\,\I_{000} \, F_2(k_1, k_2;k_3)\, P_L(k_1)\,P_L(k_2)+2~{\rm perm.} 
+ \nn\\ & 
+ 2\,b_1^2\,f\,(\I_{200}+\I_{020})\,F_2(k_1, k_2; k_3)\,P_L(k_1)\,P_L(k_2)+2~{\rm perm.}  
+ \nn \\ & 
+ 2\,b_1\,f^2\,\I_{220}\,F_2(k_1, k_2; k_3)\,P_L(k_1)\,P_L(k_2)+2~{\rm perm.} 
\,, \\
\label{e:B0B2}
B_{0}^{(b2)} & = 
b_1^2\, b_2\,\I_{000}\,P_L(k_1)\,P_L(k_2)+2~{\rm perm.}
+ \nn \\ & 
+ b_1\, b_2\, f\,(\I_{200}+\I_{020})\, P_L(k_1) \, P_L(k_2) + 2~{\rm perm.}
+ \nn \\ &
+ b_2\, f^2\, \I_{220}P_L(k_1) P_L(k_2) +2~{\rm perm.} 
\,, \\ 
\label{e:B0S2}
B_{0}^{(S_2)} & = 
2\,b_1^2\, \bGt\, \I_{000} \, S(k_1, k_2; k_3)\,P_L(k_1)\, P_L(k_2)+2~{\rm perm.}  
+ \nn \\ & 
+ 2\,b_1\, \bGt\, f\,(\I_{200}+ \I_{020})\, S(k_1, k_2; k_3)\, P_L(k_1)\, P_L(k_2)+2~{\rm perm.} 
+ \nn \\ & 
+ 2\,\bGt\, f^2\, \I_{220}\, S(k_1, k_2; k_3)\, P_L(k_1)\, P_L(k_2) + 2~{\rm perm.} 
\,, \\
\label{e:B0G2}
B_{0}^{(G_2)} & = 
2\,b_1^2\, f\, \I_{002}\, G_2(k_1, k_2; k_3)\, P_L(k_1)\, P_L(k_2) + 2~{\rm perm.} 
+ \nn \\ & 
+ 2\, b_1\, f^2\, (\I_{202} + \I_{022})\, G_2(k_1, k_2; k_3)\, P_L(k_1)\, P_L(k_2) + 2~{\rm perm.}  
+ \nn \\ & 
+ 2\, f^3\, \I_{222}\, G_2(k_1, k_2; k_3)\, P_L(k_1)\, P_L(k_2)+ 2~{\rm perm.} 
\,, \\
\label{e:B0Mixed}
B_{0}^{(\text{mixed})} & = 
- b_1^3\, f\, \Big(\frac{k_3}{k_1}\I_{101}+\frac{k_3}{k_2}\I_{011}\Big)\, P_L(k_1)\, P_L(k_2) +2~{\rm perm.} 
- \nn \\ & 
-b_1^2\, f^2\, \Big[\frac{k_3}{k_1}(\I_{301} + 2\I_{121})+\frac{k_3}{k_2}(\I_{031}+2\I_{211})\Big]\, P_L(k_1)\, P_L(k_2) + 2~{\rm perm.}
- \nn \\ & 
- b_1\, f^3\, \Big[\frac{k_3}{k_1}(\I_{141}+2 \I_{321})+ \frac{k_3}{k_2}(\I_{411}+2\I_{231})\Big]\, P_L(k_1)\, P_L(k_2) + 2~{\rm perm.}
- \nn \\ & 
- f^4\, \Big(\frac{k_3}{k_1}\I_{341} + \frac{k_3}{k_2}\I_{431}\Big)\, P_L(k_1)\, P_L(k_2)+2~{\rm perm.}
\,.
\end{align}
Here the permutations are intended to apply as well on the $\mathcal{I}_{\alpha \beta \gamma} $ integrals. The shot-noise contribution to the monopole is given by
\begin{align}
B_{0}^{(\text{shot-noise})} & =  
\frac{1}{\tilde{n}}\Big\{  b_1^2\,\I_{000}\,\Big[P_L(k_1)+ P_L(k_2) + P_L(k_3)\Big] +
\nn \\ &  
+ b_1 f\Big[\I_{200}\, P(k_1) + \I_{020}\, P(k_2) +\I_{002}\, P(k_3) \Big] + \nn \\ &  
+ f^2\Big[\I_{400}P(k_1) + \I_{040}P(k_2)P +\I_{004}P(k_3) \Big]\Big\} + 
\frac{\I_{000}}{\bar{n}^2}
\end{align}

Then the quadrupole and hexadecapole, defined as 
\begin{align}
B_2(k_1, k_2, k_3) &= \int_{-1}^{+1}d\mu_1\int_0^{2\pi}d\xi B(k_1, k_2, k_2, \mu_1, \xi)\Big[\frac{1}{4}\sqrt{\frac{5}{\pi}}(3\mu_1^2 - 1)\Big]
\\
B_4(k_1, k_2, k_3) &= \int_{-1}^{+1}d\mu_1\int_0^{2\pi}d\xi B(k_1, k_2, k_2, \mu_1, \xi)\Big[\frac{3}{16}\sqrt{\frac{1}{\pi}}(35\mu_1^4 - 30\mu_1^2 + 3)\Big]
\end{align}
 can be readily written starting from eq.s~(\ref{e:B0F2})-(\ref{e:B0Mixed}), replacing the $\mathcal{I}_{\alpha \beta \gamma}$ integrals with the quantities
\be
\J^{(2)}_{\alpha\beta\gamma}  \equiv \frac14\sqrt{\frac{5}{\pi}}\left(3\,\I_{\alpha+2\,\beta\,\gamma}-\I_{\alpha\,\beta\,\gamma}\right)\,,
\ee
in the quadrupole case and with
\be
\J^{(4)}_{\alpha\beta\gamma}  \equiv \frac3{16\,\sqrt{\pi}}\left(35\,\I_{\alpha+4\,\beta\,\gamma}-30\I_{\alpha+2\,\beta\,\gamma}+3\I_{\alpha\,\beta\,\gamma}\right)\,,
\ee
for the hexadecapole.

\bibliographystyle{JHEP}
\bibliography{cosmologia}

\providecommand{\href}[2]{#2}\begingroup\raggedright\begin{thebibliography}{10}

\bibitem{LaureijsEtal2011}
R.~{Laureijs}, J.~{Amiaux}, S.~{Arduini}, J.~. {Augu{\`e}res}, J.~{Brinchmann},
  R.~{Cole}, M.~{Cropper}, C.~{Dabin}, L.~{Duvet}, A.~{Ealet}, and et~al., {\it
  {Euclid Definition Study Report}},  {\em ArXiv: 1110.3193} (Oct., 2011)
  [\href{http://xxx.lanl.gov/abs/1110.3193}{{\tt arXiv:1110.3193}}].

\bibitem{LeviEtal2013}
M.~{Levi}, C.~{Bebek}, T.~{Beers}, R.~{Blum}, R.~{Cahn}, D.~{Eisenstein},
  B.~{Flaugher}, K.~{Honscheid}, R.~{Kron}, O.~{Lahav}, P.~{McDonald},
  N.~{Roe}, D.~{Schlegel}, and {representing the DESI collaboration}, {\it {The
  DESI Experiment, a whitepaper for Snowmass 2013}},  {\em ArXiv: 1308.0847}
  (Aug., 2013) [\href{http://xxx.lanl.gov/abs/1308.0847}{{\tt
  arXiv:1308.0847}}].

\bibitem{DoreEtal2014A}
O.~{Dor{\'e}}, J.~{Bock}, M.~{Ashby}, P.~{Capak}, A.~{Cooray}, R.~{de Putter},
  T.~{Eifler}, N.~{Flagey}, Y.~{Gong}, S.~{Habib}, K.~{Heitmann}, C.~{Hirata},
  W.-S. {Jeong}, R.~{Katti}, P.~{Korngut}, E.~{Krause}, D.-H. {Lee},
  D.~{Masters}, P.~{Mauskopf}, G.~{Melnick}, B.~{Mennesson}, H.~{Nguyen},
  K.~{{\"O}berg}, A.~{Pullen}, A.~{Raccanelli}, R.~{Smith}, Y.-S. {Song},
  V.~{Tolls}, S.~{Unwin}, T.~{Venumadhav}, M.~{Viero}, M.~{Werner}, and
  M.~{Zemcov}, {\it {Cosmology with the SPHEREX All-Sky Spectral Survey}},
  {\em ArXiv e-prints} (Dec., 2014)
  [\href{http://xxx.lanl.gov/abs/1412.4872}{{\tt arXiv:1412.4872}}].

\bibitem{SlepianEtal2017}
Z.~{Slepian}, D.~J. {Eisenstein}, J.~R. {Brownstein}, C.-H. {Chuang},
  H.~{Gil-Mar{\'{\i}}n}, S.~{Ho}, F.-S. {Kitaura}, W.~J. {Percival}, A.~J.
  {Ross}, G.~{Rossi}, H.-J. {Seo}, A.~{Slosar}, and M.~{Vargas-Maga{\~n}a},
  {\it {Detection of baryon acoustic oscillation features in the large-scale
  three-point correlation function of SDSS BOSS DR12 CMASS galaxies}},  {\em
  \mnras} {\bf 469} (Aug., 2017) 1738--1751,
  [\href{http://xxx.lanl.gov/abs/1607.0609}{{\tt arXiv:1607.0609}}].

\bibitem{VeropalumboEtal2021}
A.~{Veropalumbo}, I.~{S{\'a}ez Casares}, E.~{Branchini}, B.~R. {Granett},
  L.~{Guzzo}, F.~{Marulli}, M.~{Moresco}, L.~{Moscardini}, A.~{Pezzotta}, and
  S.~{de la Torre}, {\it {A joint 2- and 3-point clustering analysis of the
  VIPERS PDR2 catalogue at z 1: breaking the degeneracy of cosmological
  parameters}},  {\em \mnras} {\bf 507} (Oct., 2021) 1184--1201,
  [\href{http://xxx.lanl.gov/abs/2106.1258}{{\tt arXiv:2106.1258}}].

\bibitem{GilMarinEtal2017}
H.~{Gil-Mar{\'{\i}}n}, W.~J. {Percival}, L.~{Verde}, J.~R. {Brownstein}, C.-H.
  {Chuang}, F.-S. {Kitaura}, S.~A. {Rodr{\'{\i}}guez-Torres}, and M.~D.
  {Olmstead}, {\it {The clustering of galaxies in the SDSS-III Baryon
  Oscillation Spectroscopic Survey: RSD measurement from the power spectrum and
  bispectrum of the DR12 BOSS galaxies}},  {\em \mnras} {\bf 465} (Feb., 2017)
  1757--1788, [\href{http://xxx.lanl.gov/abs/1606.0043}{{\tt
  arXiv:1606.0043}}].

\bibitem{PearsonSamushia2018}
D.~W. {Pearson} and L.~{Samushia}, {\it {A Detection of the Baryon Acoustic
  Oscillation features in the SDSS BOSS DR12 Galaxy Bispectrum}},  {\em \mnras}
  {\bf 478} (Aug., 2018) 4500--4512,
  [\href{http://xxx.lanl.gov/abs/1712.0497}{{\tt arXiv:1712.0497}}].

\bibitem{DAmicoEtal2020}
G.~{d'Amico}, J.~{Gleyzes}, N.~{Kokron}, K.~{Markovic}, L.~{Senatore},
  P.~{Zhang}, F.~{Beutler}, and H.~{Gil-Mar{\'\i}n}, {\it {The cosmological
  analysis of the SDSS/BOSS data from the Effective Field Theory of Large-Scale
  Structure}},  {\em \jcap} {\bf 2020} (May, 2020) 005,
  [\href{http://xxx.lanl.gov/abs/1909.0527}{{\tt arXiv:1909.0527}}].

\bibitem{PhilcoxIvanov2022}
O.~H.~E. {Philcox} and M.~M. {Ivanov}, {\it {BOSS DR12 full-shape cosmology:
  {\ensuremath{\Lambda}} CDM constraints from the large-scale galaxy power
  spectrum and bispectrum monopole}},  {\em \prd} {\bf 105} (Feb., 2022)
  043517, [\href{http://xxx.lanl.gov/abs/2112.0451}{{\tt arXiv:2112.0451}}].

\bibitem{DAmicoEtal2022A}
G.~{D'Amico}, M.~{Lewandowski}, L.~{Senatore}, and P.~{Zhang}, {\it {Limits on
  primordial non-Gaussianities from BOSS galaxy-clustering data}},  {\em arXiv
  e-prints} (Jan., 2022) arXiv:2201.11518,
  [\href{http://xxx.lanl.gov/abs/2201.1151}{{\tt arXiv:2201.1151}}].

\bibitem{CabassEtal2022A}
G.~{Cabass}, M.~M. {Ivanov}, O.~H.~E. {Philcox}, M.~{Simonovi{\'c}}, and
  M.~{Zaldarriaga}, {\it {Constraints on Single-Field Inflation from the BOSS
  Galaxy Survey}},  {\em arXiv e-prints} (Jan., 2022) arXiv:2201.07238,
  [\href{http://xxx.lanl.gov/abs/2201.0723}{{\tt arXiv:2201.0723}}].

\bibitem{CabassEtal2022B}
G.~{Cabass}, M.~M. {Ivanov}, O.~H.~E. {Philcox}, M.~{Simonovi{\'c}}, and
  M.~{Zaldarriaga}, {\it {Constraints on Multi-Field Inflation from the BOSS
  Galaxy Survey}},  {\em arXiv e-prints} (Apr., 2022) arXiv:2204.01781,
  [\href{http://xxx.lanl.gov/abs/2204.0178}{{\tt arXiv:2204.0178}}].

\bibitem{AlamEtal2017}
S.~{Alam}, M.~{Ata}, S.~{Bailey}, F.~{Beutler}, D.~{Bizyaev}, J.~A. {Blazek},
  A.~S. {Bolton}, J.~R. {Brownstein}, A.~{Burden}, C.-H. {Chuang},
  J.~{Comparat}, A.~J. {Cuesta}, K.~S. {Dawson}, D.~J. {Eisenstein},
  S.~{Escoffier}, H.~{Gil-Mar{\'{\i}}n}, J.~N. {Grieb}, N.~{Hand}, S.~{Ho},
  K.~{Kinemuchi}, D.~{Kirkby}, F.~{Kitaura}, E.~{Malanushenko},
  V.~{Malanushenko}, C.~{Maraston}, C.~K. {McBride}, R.~C. {Nichol}, M.~D.
  {Olmstead}, D.~{Oravetz}, N.~{Padmanabhan}, N.~{Palanque-Delabrouille},
  K.~{Pan}, M.~{Pellejero-Ibanez}, W.~J. {Percival}, P.~{Petitjean},
  F.~{Prada}, A.~M. {Price-Whelan}, B.~A. {Reid}, S.~A.
  {Rodr{\'{\i}}guez-Torres}, N.~A. {Roe}, A.~J. {Ross}, N.~P. {Ross},
  G.~{Rossi}, J.~A. {Rubi{\~n}o-Mart{\'{\i}}n}, S.~{Saito},
  S.~{Salazar-Albornoz}, L.~{Samushia}, A.~G. {S{\'a}nchez}, S.~{Satpathy},
  D.~J. {Schlegel}, D.~P. {Schneider}, C.~G. {Sc{\'o}ccola}, H.-J. {Seo}, E.~S.
  {Sheldon}, A.~{Simmons}, A.~{Slosar}, M.~A. {Strauss}, M.~E.~C. {Swanson},
  D.~{Thomas}, J.~L. {Tinker}, R.~{Tojeiro}, M.~V. {Maga{\~n}a}, J.~A.
  {Vazquez}, L.~{Verde}, D.~A. {Wake}, Y.~{Wang}, D.~H. {Weinberg}, M.~{White},
  W.~M. {Wood-Vasey}, C.~{Y{\`e}che}, I.~{Zehavi}, Z.~{Zhai}, and G.-B. {Zhao},
  {\it {The clustering of galaxies in the completed SDSS-III Baryon Oscillation
  Spectroscopic Survey: cosmological analysis of the DR12 galaxy sample}},
  {\em \mnras} {\bf 470} (Sept., 2017) 2617--2652,
  [\href{http://xxx.lanl.gov/abs/1607.0315}{{\tt arXiv:1607.0315}}].

\bibitem{RuggeriEtal2019}
R.~{Ruggeri}, W.~J. {Percival}, H.~{Gil-Mar{\'{\i}}n}, F.~{Beutler}, E.-M.
  {Mueller}, F.~{Zhu}, N.~{Padmanabhan}, G.-B. {Zhao}, P.~{Zarrouk}, A.~G.
  {S{\'a}nchez}, J.~{Bautista}, J.~{Brinkmann}, J.~R. {Brownstein},
  F.~{Baumgarten}, C.-H. {Chuang}, K.~{Dawson}, H.-J. {Seo}, R.~{Tojeiro}, and
  C.~{Zhao}, {\it {The clustering of the SDSS-IV extended Baryon Oscillation
  Spectroscopic Survey DR14 quasar sample: measuring the evolution of the
  growth rate using redshift-space distortions between redshift 0.8 and 2.2}},
  {\em \mnras} {\bf 483} (Mar., 2019) 3878--3887,
  [\href{http://xxx.lanl.gov/abs/1801.0289}{{\tt arXiv:1801.0289}}].

\bibitem{SongTaruyaOka2015}
Y.-S. {Song}, A.~{Taruya}, and A.~{Oka}, {\it {Cosmology with anisotropic
  galaxy clustering from the combination of power spectrum and bispectrum}},
  {\em \jcap} {\bf 8} (Aug., 2015) 007,
  [\href{http://xxx.lanl.gov/abs/1502.0309}{{\tt arXiv:1502.0309}}].

\bibitem{GagraniSamushia2017}
P.~{Gagrani} and L.~{Samushia}, {\it {Information Content of the Angular
  Multipoles of Redshift-Space Galaxy Bispectrum}},  {\em \mnras} {\bf 467}
  (May, 2017) 928--935, [\href{http://xxx.lanl.gov/abs/1610.0348}{{\tt
  arXiv:1610.0348}}].

\bibitem{YankelevichPorciani2019}
V.~{Yankelevich} and C.~{Porciani}, {\it {Cosmological information in the
  redshift-space bispectrum}},  {\em \mnras} (Nov., 2018)
  [\href{http://xxx.lanl.gov/abs/1807.0707}{{\tt arXiv:1807.0707}}].

\bibitem{ChudaykinIvanov2019}
A.~{Chudaykin} and M.~M. {Ivanov}, {\it {Measuring neutrino masses with
  large-scale structure: Euclid forecast with controlled theoretical error}},
  {\em \jcap} {\bf 2019} (Nov., 2019) 034,
  [\href{http://xxx.lanl.gov/abs/1907.0666}{{\tt arXiv:1907.0666}}].

\bibitem{HahnEtal2020}
C.~{Hahn}, F.~{Villaescusa-Navarro}, E.~{Castorina}, and R.~{Scoccimarro}, {\it
  {Constraining M$_{{\ensuremath{\nu}}}$ with the bispectrum. Part I. Breaking
  parameter degeneracies}},  {\em \jcap} {\bf 2020} (Mar., 2020) 040,
  [\href{http://xxx.lanl.gov/abs/1909.1110}{{\tt arXiv:1909.1110}}].

\bibitem{GualdiVerde2020}
D.~{Gualdi} and L.~{Verde}, {\it {Galaxy redshift-space bispectrum: the
  importance of being anisotropic}},  {\em \jcap} {\bf 2020} (June, 2020) 041,
  [\href{http://xxx.lanl.gov/abs/2003.1207}{{\tt arXiv:2003.1207}}].

\bibitem{HahnVillaescusaNavarro2021}
C.~{Hahn} and F.~{Villaescusa-Navarro}, {\it {Constraining
  M$_{{\ensuremath{\nu}}}$ with the bispectrum. Part II. The information
  content of the galaxy bispectrum monopole}},  {\em \jcap} {\bf 2021} (Apr.,
  2021) 029, [\href{http://xxx.lanl.gov/abs/2012.0220}{{\tt arXiv:2012.0220}}].

\bibitem{AgarwalEtal2021}
N.~{Agarwal}, V.~{Desjacques}, D.~{Jeong}, and F.~{Schmidt}, {\it {Information
  content in the redshift-space galaxy power spectrum and bispectrum}},  {\em
  \jcap} {\bf 2021} (Mar., 2021) 021,
  [\href{http://xxx.lanl.gov/abs/2007.0434}{{\tt arXiv:2007.0434}}].

\bibitem{IvanovEtal2021A}
M.~M. {Ivanov}, O.~H.~E. {Philcox}, T.~{Nishimichi}, M.~{Simonovi{\'c}},
  M.~{Takada}, and M.~{Zaldarriaga}, {\it {Precision analysis of the
  redshift-space galaxy bispectrum}},  {\em arXiv e-prints} (Oct., 2021)
  arXiv:2110.10161, [\href{http://xxx.lanl.gov/abs/2110.1016}{{\tt
  arXiv:2110.1016}}].

\bibitem{HivonEtal1995}
E.~{Hivon}, F.~R. {Bouchet}, S.~{Colombi}, and R.~{Juszkiewicz}, {\it {Redshift
  distortions of clustering: a Lagrangian approach.}},  {\em \aap} {\bf 298}
  (June, 1995) 643, [\href{http://xxx.lanl.gov/abs/astro-ph/9407049}{{\tt
  astro-ph/9407049}}].

\bibitem{BernardeauEtal2002}
F.~Bernardeau, S.~Colombi, E.~Gazta{\~n}aga, and R.~Scoccimarro, {\it
  Large-scale structure of the universe and cosmological perturbation theory},
  {\em \physrep} {\bf 367} (Sept., 2002) 1--3,
  [\href{http://xxx.lanl.gov/abs/astro-ph/0112551}{{\tt astro-ph/0112551}}].

\bibitem{VerdeEtal1998}
L.~Verde, A.~F. Heavens, S.~Matarrese, and L.~Moscardini, {\it Large-scale bias
  in the universe - ii. redshift-space bispectrum},  {\em \mnras} {\bf 300}
  (Nov., 1998) 747--756, [\href{http://xxx.lanl.gov/abs/astro-ph/}{{\tt
  astro-ph/}}].

\bibitem{ScoccimarroCouchmanFrieman1999}
R.~Scoccimarro, H.~M.~P. Couchman, and J.~A. Frieman, {\it The bispectrum as a
  signature of gravitational instability in redshift space},  {\em \apj} {\bf
  517} (June, 1999) 531--540,
  [\href{http://xxx.lanl.gov/abs/astro-ph/9808305}{{\tt astro-ph/9808305}}].

\bibitem{Scoccimarro2000B}
R.~Scoccimarro, {\it The bispectrum: From theory to observations},  {\em \apj}
  {\bf 544} (Dec., 2000) 597--615,
  [\href{http://xxx.lanl.gov/abs/astro-ph/0004086}{{\tt astro-ph/0004086}}].

\bibitem{GilMarinEtal2015}
H.~{Gil-Mar{\'{\i}}n}, J.~{Nore{\~n}a}, L.~{Verde}, W.~J. {Percival},
  C.~{Wagner}, M.~{Manera}, and D.~P. {Schneider}, {\it {The power spectrum and
  bispectrum of SDSS DR11 BOSS galaxies - I. Bias and gravity}},  {\em \mnras}
  {\bf 451} (July, 2015) 539--580,
  [\href{http://xxx.lanl.gov/abs/1407.5668}{{\tt arXiv:1407.5668}}].

\bibitem{GilMarinEtal2015B}
H.~{Gil-Mar{\'{\i}}n}, L.~{Verde}, J.~{Nore{\~n}a}, A.~J. {Cuesta},
  L.~{Samushia}, W.~J. {Percival}, C.~{Wagner}, M.~{Manera}, and D.~P.
  {Schneider}, {\it {The power spectrum and bispectrum of SDSS DR11 BOSS
  galaxies - II. Cosmological interpretation}},  {\em \mnras} {\bf 452} (Sept.,
  2015) 1914--1921, [\href{http://xxx.lanl.gov/abs/1408.0027}{{\tt
  arXiv:1408.0027}}].

\bibitem{GilMarinEtal2014}
H.~{Gil-Mar{\'{\i}}n}, C.~{Wagner}, J.~{Nore{\~n}a}, L.~{Verde}, and
  W.~{Percival}, {\it {Dark matter and halo bispectrum in redshift space:
  theory and applications}},  {\em \jcap} {\bf 12} (Dec., 2014) 029,
  [\href{http://xxx.lanl.gov/abs/1407.1836}{{\tt arXiv:1407.1836}}].

\bibitem{GualdiGilMarinVerde2021}
D.~{Gualdi}, H.~{Gil-Mar{\'\i}n}, and L.~{Verde}, {\it {Joint analysis of
  anisotropic power spectrum, bispectrum and trispectrum: application to N-body
  simulations}},  {\em \jcap} {\bf 2021} (July, 2021) 008,
  [\href{http://xxx.lanl.gov/abs/2104.0397}{{\tt arXiv:2104.0397}}].

\bibitem{HashimotoRaseraTaruya2017}
I.~{Hashimoto}, Y.~{Rasera}, and A.~{Taruya}, {\it {Precision cosmology with
  redshift-space bispectrum: A perturbation theory based model at one-loop
  order}},  {\em \prd} {\bf 96} (Aug., 2017) 043526,
  [\href{http://xxx.lanl.gov/abs/1705.0257}{{\tt arXiv:1705.0257}}].

\bibitem{TaruyaNishimichiSaito2010}
A.~{Taruya}, T.~{Nishimichi}, and S.~{Saito}, {\it {Baryon acoustic
  oscillations in 2D: Modeling redshift-space power spectrum from perturbation
  theory}},  {\em \prd} {\bf 82} (Sept., 2010) 063522,
  [\href{http://xxx.lanl.gov/abs/1006.0699}{{\tt arXiv:1006.0699}}].

\bibitem{NishimichiEtal2020}
T.~{Nishimichi}, G.~{D'Amico}, M.~M. {Ivanov}, L.~{Senatore},
  M.~{Simonovi{\'c}}, M.~{Takada}, M.~{Zaldarriaga}, and P.~{Zhang}, {\it
  {Blinded challenge for precision cosmology with large-scale structure:
  Results from effective field theory for the redshift-space galaxy power
  spectrum}},  {\em \prd} {\bf 102} (Dec., 2020) 123541,
  [\href{http://xxx.lanl.gov/abs/2003.0827}{{\tt arXiv:2003.0827}}].

\bibitem{OddoEtal2020}
A.~{Oddo}, E.~{Sefusatti}, C.~{Porciani}, P.~{Monaco}, and A.~G. {S{\'a}nchez},
  {\it {Toward a robust inference method for the galaxy bispectrum: likelihood
  function and model selection}},  {\em \jcap} {\bf 2020} (Mar., 2020) 056,
  [\href{http://xxx.lanl.gov/abs/1908.0177}{{\tt arXiv:1908.0177}}].

\bibitem{AlkhanishviliEtal2021}
D.~{Alkhanishvili}, C.~{Porciani}, E.~{Sefusatti}, M.~{Biagetti}, A.~{Lazanu},
  A.~{Oddo}, and V.~{Yankelevich}, {\it {The reach of next-to-leading-order
  perturbation theory for the matter bispectrum}},  {\em arXiv e-prints} (July,
  2021) arXiv:2107.08054, [\href{http://xxx.lanl.gov/abs/2107.0805}{{\tt
  arXiv:2107.0805}}].

\bibitem{OddoEtal2021}
A.~{Oddo}, F.~{Rizzo}, E.~{Sefusatti}, C.~{Porciani}, and P.~{Monaco}, {\it
  {Cosmological parameters from the likelihood analysis of the galaxy power
  spectrum and bispectrum in real space}},  {\em \jcap} {\bf 2021} (Nov., 2021)
  038, [\href{http://xxx.lanl.gov/abs/2108.0320}{{\tt arXiv:2108.0320}}].

\bibitem{FryGaztanaga1993}
J.~N. Fry and E.~Gazta{\~n}aga, {\it Biasing and hierarchical statistics in
  large-scale structure},  {\em \apj} {\bf 413} (Aug., 1993) 447--452,
  [\href{http://xxx.lanl.gov/abs/astro-ph/9302009}{{\tt astro-ph/9302009}}].

\bibitem{ChanScoccimarroSheth2012}
K.~C. Chan, R.~Scoccimarro, and R.~K. Sheth, {\it {Gravity and large-scale
  nonlocal bias}},  {\em \prd} {\bf 85} (Apr., 2012) 083509,
  [\href{http://xxx.lanl.gov/abs/1201.3614}{{\tt arXiv:1201.3614}}].

\bibitem{BaldaufEtal2012}
T.~{Baldauf}, U.~{Seljak}, V.~{Desjacques}, and P.~{McDonald}, {\it {Evidence
  for quadratic tidal tensor bias from the halo bispectrum}},  {\em \prd} {\bf
  86} (Oct., 2012) 083540, [\href{http://xxx.lanl.gov/abs/1201.4827}{{\tt
  arXiv:1201.4827}}].

\bibitem{DesjacquesJeongSchmidt2018B}
V.~{Desjacques}, D.~{Jeong}, and F.~{Schmidt}, {\it {The galaxy power spectrum
  and bispectrum in redshift space}},  {\em \jcap} {\bf 2018} (Dec., 2018) 035,
  [\href{http://xxx.lanl.gov/abs/1806.0401}{{\tt arXiv:1806.0401}}].

\bibitem{Schmidt2016}
F.~{Schmidt}, {\it {Towards a self-consistent halo model for the nonlinear
  large-scale structure}},  {\em \prd} {\bf 93} (Mar., 2016) 063512,
  [\href{http://xxx.lanl.gov/abs/1511.0223}{{\tt arXiv:1511.0223}}].

\bibitem{PerkoEtal2016A}
A.~{Perko}, L.~{Senatore}, E.~{Jennings}, and R.~H. {Wechsler}, {\it {Biased
  Tracers in Redshift Space in the EFT of Large-Scale Structure}},  {\em ArXiv
  e-prints} (Oct., 2016) [\href{http://xxx.lanl.gov/abs/1610.0932}{{\tt
  arXiv:1610.0932}}].

\bibitem{GinzburgDesjacquesChan2017}
D.~{Ginzburg}, V.~{Desjacques}, and K.~C. {Chan}, {\it {Shot noise and biased
  tracers: A new look at the halo model}},  {\em \prd} {\bf 96} (Oct., 2017)
  083528, [\href{http://xxx.lanl.gov/abs/1706.0873}{{\tt arXiv:1706.0873}}].

\bibitem{Scoccimarro2015}
R.~{Scoccimarro}, {\it {Fast estimators for redshift-space clustering}},  {\em
  \prd} {\bf 92} (Oct., 2015) 083532,
  [\href{http://xxx.lanl.gov/abs/1506.0272}{{\tt arXiv:1506.0272}}].

\bibitem{ByunKrause2022}
J.~{Byun} and E.~{Krause}, {\it {Modal compression of the redshift-space galaxy
  bispectrum}},  {\em arXiv e-prints} (May, 2022) arXiv:2205.04579,
  [\href{http://xxx.lanl.gov/abs/2205.0457}{{\tt arXiv:2205.0457}}].

\bibitem{GriebEtal2016}
J.~N. Grieb, A.~G. S{\'a}nchez, S.~Salazar-Albornoz, and C.~Dalla~Vecchia, {\it
  {Gaussian covariance matrices for anisotropic galaxy clustering
  measurements}},  {\em \mnras} {\bf 457} (Apr., 2016) 1577--1592,
  [\href{http://xxx.lanl.gov/abs/1509.0429}{{\tt arXiv:1509.0429}}].

\bibitem{SefusattiEtal2016}
E.~{Sefusatti}, M.~{Crocce}, R.~{Scoccimarro}, and H.~M.~P. {Couchman}, {\it
  {Accurate estimators of correlation functions in Fourier space}},  {\em
  \mnras} {\bf 460} (Aug., 2016) 3624--3636,
  [\href{http://xxx.lanl.gov/abs/1512.0729}{{\tt arXiv:1512.0729}}].

\bibitem{MonacoTheunsTaffoni2002}
P.~Monaco, T.~{Theuns}, and G.~{Taffoni}, {\it {The pinocchio algorithm:
  pinpointing orbit-crossing collapsed hierarchical objects in a linear density
  field}},  {\em \mnras} {\bf 331} (Apr., 2002) 587--608,
  [\href{http://xxx.lanl.gov/abs/astro-ph/}{{\tt astro-ph/}}].

\bibitem{MonacoEtal2013}
P.~Monaco, E.~Sefusatti, S.~Borgani, M.~Crocce, P.~Fosalba, R.~K. Sheth, and
  T.~Theuns, {\it {An accurate tool for the fast generation of dark matter halo
  catalogues}},  {\em \mnras} {\bf 433} (Aug., 2013) 2389--2402,
  [\href{http://xxx.lanl.gov/abs/1305.1505}{{\tt arXiv:1305.1505}}].

\bibitem{MunariEtal2017}
E.~{Munari}, P.~{Monaco}, E.~{Sefusatti}, E.~{Castorina}, F.~G. {Mohammad},
  S.~{Anselmi}, and S.~{Borgani}, {\it {Improving fast generation of halo
  catalogues with higher order Lagrangian perturbation theory}},  {\em \mnras}
  {\bf 465} (Mar., 2017) 4658--4677,
  [\href{http://xxx.lanl.gov/abs/1605.0478}{{\tt arXiv:1605.0478}}].

\bibitem{Barreira2019}
A.~{Barreira}, {\it {The squeezed matter bispectrum covariance with
  responses}},  {\em \jcap} {\bf 2019} (Mar., 2019) 008,
  [\href{http://xxx.lanl.gov/abs/1901.0124}{{\tt arXiv:1901.0124}}].

\bibitem{BiagettiEtal2021A}
M.~{Biagetti}, L.~{Castiblanco}, J.~{Nore{\~n}a}, and E.~{Sefusatti}, {\it {The
  Covariance of Squeezed Bispectrum Configurations}},  {\em arXiv e-prints}
  (Nov., 2021) arXiv:2111.05887, [\href{http://xxx.lanl.gov/abs/2111.0588}{{\tt
  arXiv:2111.0588}}].

\bibitem{SellentinHeavens2016}
E.~{Sellentin} and A.~F. {Heavens}, {\it {Parameter inference with estimated
  covariance matrices}},  {\em \mnras} {\bf 456} (Feb., 2016) L132--L136,
  [\href{http://xxx.lanl.gov/abs/1511.0596}{{\tt arXiv:1511.0596}}].

\bibitem{Anderson2003}
T.~W. Anderson, {\em An introduction to multivariate statistical analysis}.
\newblock Wiley New York, 1958.

\bibitem{HartlapEtal2009}
J.~{Hartlap}, T.~{Schrabback}, P.~{Simon}, and P.~{Schneider}, {\it {The
  non-Gaussianity of the cosmic shear likelihood or how odd is the Chandra Deep
  Field South?}},  {\em \aap} {\bf 504} (Sept., 2009) 689--703,
  [\href{http://xxx.lanl.gov/abs/0901.3269}{{\tt arXiv:0901.3269}}].

\bibitem{GualdiEtal2020}
D.~{Gualdi}, H.~{Gil-Mar{\'\i}n}, M.~{Manera}, B.~{Joachimi}, and O.~{Lahav},
  {\it {GEOMAX: beyond linear compression for three-point galaxy clustering
  statistics}},  {\em \mnras} {\bf 497} (July, 2020) 776--792,
  [\href{http://xxx.lanl.gov/abs/1912.0101}{{\tt arXiv:1912.0101}}].

\bibitem{Peebles1980}
P.~J.~E. {Peebles}, {\em The large-scale structure of the universe}.
\newblock Princeton, N.J., Princeton University Press, 1980.~435 p., 1980.

\bibitem{MatarreseVerdeHeavens1997}
S.~Matarrese, L.~Verde, and A.~F. Heavens, {\it Large-scale bias in the
  universe: bispectrum method},  {\em \mnras} {\bf 290} (Oct., 1997) 651--662,
  [\href{http://xxx.lanl.gov/abs/astro-ph/9706059}{{\tt astro-ph/9706059}}].

\bibitem{SugiyamaEtal2019}
N.~S. {Sugiyama}, S.~{Saito}, F.~{Beutler}, and H.-J. {Seo}, {\it {A complete
  FFT-based decomposition formalism for the redshift-space bispectrum}},  {\em
  \mnras} {\bf 484} (Mar., 2019) 364--384,
  [\href{http://xxx.lanl.gov/abs/1803.0213}{{\tt arXiv:1803.0213}}].

\bibitem{ShethChanScoccimarro2013}
R.~K. {Sheth}, K.~C. {Chan}, and R.~{Scoccimarro}, {\it {Nonlocal Lagrangian
  bias}},  {\em \prd} {\bf 87} (Apr., 2013) 083002,
  [\href{http://xxx.lanl.gov/abs/1207.7117}{{\tt arXiv:1207.7117}}].

\bibitem{LazeyrasEtal2016}
T.~{Lazeyras}, C.~Wagner, T.~Baldauf, and F.~Schmidt, {\it {Precision
  measurement of the local bias of dark matter halos}},  {\em \jcap} {\bf 2}
  (Feb., 2016) 018, [\href{http://xxx.lanl.gov/abs/1511.0109}{{\tt
  arXiv:1511.0109}}].

\bibitem{LazeyrasSchmidt2018}
T.~{Lazeyras} and F.~{Schmidt}, {\it {Beyond LIMD bias: a measurement of the
  complete set of third-order halo bias parameters}},  {\em \jcap} {\bf 9}
  (Sept., 2018) 008, [\href{http://xxx.lanl.gov/abs/1712.0753}{{\tt
  arXiv:1712.0753}}].

\bibitem{EggemeierEtal2020}
A.~{Eggemeier}, R.~{Scoccimarro}, M.~{Crocce}, A.~{Pezzotta}, and A.~G.
  {S{\'a}nchez}, {\it {Testing one-loop galaxy bias: Power spectrum}},  {\em
  \prd} {\bf 102} (Nov., 2020) 103530,
  [\href{http://xxx.lanl.gov/abs/2006.0972}{{\tt arXiv:2006.0972}}].

\bibitem{EggemeierEtal2021}
A.~{Eggemeier}, R.~{Scoccimarro}, R.~E. {Smith}, M.~{Crocce}, A.~{Pezzotta},
  and A.~G. {S{\'a}nchez}, {\it {Testing one-loop galaxy bias: Joint analysis
  of power spectrum and bispectrum}},  {\em \prd} {\bf 103} (June, 2021)
  123550, [\href{http://xxx.lanl.gov/abs/2102.0690}{{\tt arXiv:2102.0690}}].

\bibitem{MorettiEtalInPrep}
C.~{Moretti}, F.~{Rizzo}, K.~{Pardede}, A.~{Oddo}, E.~{Sefusatti},
  C.~{Porciani}, and P.~{Monaco} in preparation.

\bibitem{ColavincenzoEtal2019}
M.~{Colavincenzo}, E.~{Sefusatti}, P.~{Monaco}, L.~{Blot}, M.~{Crocce},
  M.~{Lippich}, A.~G. {S{\'a}nchez}, M.~A. {Alvarez}, A.~{Agrawal}, S.~{Avila},
  A.~{Balaguera-Antol{\'{\i}}nez}, R.~{Bond}, S.~{Codis}, C.~{Dalla Vecchia},
  A.~{Dorta}, P.~{Fosalba}, A.~{Izard}, F.-S. {Kitaura}, M.~{Pellejero-Ibanez},
  G.~{Stein}, M.~{Vakili}, and G.~{Yepes}, {\it {Comparing approximate methods
  for mock catalogues and covariance matrices - III: bispectrum}},  {\em
  \mnras} {\bf 482} (Feb., 2019) 4883--4905,
  [\href{http://xxx.lanl.gov/abs/1806.0949}{{\tt arXiv:1806.0949}}].

\bibitem{DesjacquesJeongSchmidt2018}
V.~{Desjacques}, D.~{Jeong}, and F.~{Schmidt}, {\it {Large-scale galaxy bias}},
   {\em \physrep} {\bf 733} (Feb., 2018) 1--193,
  [\href{http://xxx.lanl.gov/abs/1611.0978}{{\tt arXiv:1611.0978}}].

\end{thebibliography}\endgroup

\end{document}